\documentclass{aa}  

\usepackage{color}
\usepackage{graphicx}
\usepackage{booktabs,tabularx}
\usepackage{txfonts}
\def\lya{Ly$\alpha$}
\def\lyb{Ly$\beta$}

\def\ha{H$\alpha$}
\def\hb{H$\beta$}

\def\oiiil{[\ion{O}{iii}]$\,\lambda5007$}

\def\oiil{[\ion{O}{ii}]$\,\lambda3727$}

\def\hi{\ion{H}{i}}
\def\hii{\ion{H}{ii}}
\def\nii{[\ion{N}{ii}]}
\def\kms{km\,s$^{-1}$}
\def\fesc{$f_\mathrm{esc}$\/(LyC)}
\def\fesclya{$f_\mathrm{esc}$\/(Ly$\alpha$)}
\def\si{\ion{Si}{ii}}
\def\nh{$N_\mathrm{HI}$}
\def\vexp{$v_\mathrm{exp}$}
\def\vlis{$v_\mathrm{LIS}$}

\def\ewha{$EW$\/(\ha)}
\def\ewhb{$EW$\/(\hb)}
\def\ewlya{$EW$\/(\lya)}
\def\ewlyai{$EW_0$\/(\lya)}
\def\ewlyab{$EW$\/(\lya)$_\mathrm{Blue}$}
\def\ewlyar{$EW$\/(\lya)$_\mathrm{Red}$}
\def\fwhmlyai{$FWHM_0$\/(\lya)}

\def\fwhmhb{$FWHM$\/(\hb)}

\begin{document}

   \title{Puzzling Lyman-alpha line profiles in green pea galaxies
}

   \author{I. Orlitov\'a \inst{\ref{inst1}}
          \and
           A. Verhamme \inst{\ref{inst2}}
          \and
           A. Henry \inst{\ref{inst3}}
          \and
           C. Scarlata \inst{\ref{inst4}}
          \and
           A. Jaskot \inst{\ref{inst5},\ref{inst8},\thanks{Hubble Fellow}}
          \and
           M. S. Oey \inst{\ref{inst6}}
          \and
           D. Schaerer \inst{\ref{inst2},\ref{inst7}}
          }

   \institute{Astronomical Institute of the Czech Academy of Sciences, 
              Bo\v cn{\'\i} II/1401, 141 00 Praha, Czech Republic
              \\
              \email{ivana.orlitova@asu.cas.cz}\label{inst1}
         \and
              Observatoire de Gen\`eve, Universit\'{e} de Gen\`{e}ve, 
              Chemin des Maillettes 51, 1290 Versoix, Switzerland \label{inst2}
         \and
             Space Telescope Science Institute, 3700 San Martin Drive, 
             Baltimore, MD 21218, USA \label{inst3} 
         \and
             Minnesota Institute for Astrophysics, University of Minnesota,
             Minneapolis, MN 55455, USA \label{inst4} 
         \and
             Department of Astronomy, Smith College, 44 College Lane, 
             Northampton, MA 01063, USA  \label{inst5}
         \and
            Department of Astronomy, University of Massachusetts -- Amherst, 
            Amherst, MA 01003, USA \label{inst8} 
         \and
             University of Michigan, Department of Astronomy,  
             311 West Hall, 1085 S. University Ave, 
             Ann Arbor, MI 48109-1107, USA \label{inst6}
         \and
            IRAP/CNRS, 14 Av. E. Belin, 31400 Toulouse, France \label{inst7}
             }

   \date{Received ; accepted }

 
  \abstract
   {The Lyman-alpha (\lya) line of hydrogen is of prime importance for 
    detecting galaxies at high redshift. For a correct data interpretation, 
    numerical radiative transfer models
    are necessary due to \lya\ resonant scattering off neutral hydrogen atoms. 
   }
   {Recent observations have discovered an escape of ionizing Lyman-continuum 
    radiation from a population of compact, actively star-forming galaxies at 
    redshift $z\sim0.2-0.3$, also known as ``green peas''.  
    For the potential similarities with high-redshift galaxies and impact 
    on the reionization of the universe, we study the green pea 
    \lya\ spectra, which are mostly double-peaked, 
    unlike in any other galaxy sample. 
    If the double peaks are a result of radiative transfer, 
    they can be a useful source of information on the green pea 
    interstellar medium and ionizing radiation escape. 
   }
   {We select a sample of twelve archival green peas and 
    we apply 
    numerical radiative transfer models to reproduce the observed \lya\ 
    spectral profiles,  using the geometry of 
    expanding, homogeneous spherical shells. 
    We use ancillary optical and ultraviolet data to constrain 
    the model parameters, 
    and we evaluate the match between the models and the observed 
    \lya\ spectra. 
    As a second step, we allow all the fitting parameters to be free, 
    and examine the agreement between the interstellar medium parameters 
    derived from the models and those from ancillary data.  
   }
   {The peculiar green pea double-peaked \lya\ line profiles 
    are not correctly reproduced by the constrained shell models.
    Conversely, unconstrained models fit the spectra, but
    parameters derived from the best-fitting models are not in agreement with 
    the ancillary data. In particular: 1) the best-fit systemic redshifts 
    are larger by 10\,--\,250\,\kms\ than those derived from optical 
    emission lines; 2) the double-peaked \lya\ profiles are best reproduced
    with low-velocity ($\lesssim\!\!150$\,\kms) outflows that contradict 
    the observed ultraviolet absorption lines of
    low-ionization-state elements with characteristic 
    velocities as large as 300\,\kms; and 3) 
    the models need to consider intrinsic \lya\ profiles 
    that are on average three times broader than the observed Balmer lines.
    }
    {Differences between the modelled and observed velocities are larger 
    for targets with prominent \lya\ blue peaks.  
    The blue peak position and flux appear to be connected to
    low column densities of neutral hydrogen, leading to \lya\ and 
    Lyman-continuum escape. This is at odds with the kinematic origin 
    of the blue peak in the homogeneous shell models. 
    Additional modelling is needed to explore  
    alternative geometries such as clumpy media and  
    non-recombination \lya\ sources to further constrain 
    the role and significance of the \lya\ double peaks. 
    }


   \maketitle
%

\section{Introduction}

Green peas (GPs) 
are a local population of compact ($<3$\,kpc), low-mass ($10^{8-10}M_\odot$)
galaxies with strong 
optical emission lines \citep{Cardamone09,Izotov11} found 
at redshifts $z<0.5$. 
Their popular name was coined by the citizen science 
{\em Galaxy Zoo} project \citep{Lintott08}, where the targets appear 
unresolved in the Sloan Digital Sky Survey (SDSS)
images and have green colours due to the 
strong \oiiil\ nebular emission line, with equivalent widths 
as large as $\sim\!1500$\,\AA.
\citet{Izotov11} have shown that green peas form a subset of luminous compact
galaxies that are present in the SDSS over a larger redshift range, 
$0.02\!\lesssim\!z\!\lesssim\!0.65$. 
Their compactness, low masses, sub-solar metallicities in the range  
12\,+\,log(O/H)\,$\sim$\,7.5\,--\,8.5, 
and \ha\
equivalent widths exceeding hundreds of \AA\ \citep{Izotov11}  
make green peas similar to high-redshift Lyman-break galaxies (LBGs) and 
Lyman-alpha emitters (LAEs). 

Green peas have recently drawn attention due to the ionizing 
Lyman continuum (LyC) radiation escape. The detected LyC that manages 
to leak into the intergalactic space proves that starburst galaxies 
could have played a role in the cosmic reionization. Such detections have
been rare to date:    
while six out of the six targeted green peas leak 
$6-46$\,\% of their LyC \citep{Izotov16,Izotov16b,Izotov18}, 
only four additional low-$z$ LyC leakers have been found over the 
past two decades, and their escape fractions are low, \fesc\,=\,1\,--\,4\% 
\citep{Bergvall06,Leitet13,Borthakur14,Leitherer16,Puschnig17}. 
At high redshift, the situation is even more challenging; numerous 
LyC leaking candidates have been refuted as lower-redshift interlopers 
\citep{Siana15}, and stringent upper limits have been set 
on the escape fractions for $z\!\sim\!1\!-\!2$ galaxies  
using the  
Galaxy Evolution Explorer (GALEX) and Hubble Space Telescope 
(HST) imaging catalogues \citep{Rutkowski16,Rutkowski17}.  
Only three convincing spectroscopic LyC detections 
\citep{deBarros16,Vanzella16,Shapley16,Vanzella18}
and one imaging detection \citep{Bian17} have recently been achieved at 
$z\!\sim\!2-4,$ reaching, however, large escape fractions, \fesc\,$>50$\%,
for each of these galaxies.

The Lyman continuum escape from green peas had been suspected due to their 
high star-formation rates, compactness, and high \oiiil/\oiil\ flux ratios,
unusual for local galaxies, which could be signatures of density-bounded
\hii\ regions \citep{Jaskot13,Nakajima13,Nakajima14,Stasinska15}.  
Furthermore, most of the green peas were known to be 
strong Lyman-alpha (\lya) emitters with unusual Lyman-alpha (\lya) 
line profiles consisting of
narrow, double-peaked emission lines, and weak ultraviolet (UV)
absorption lines of 
low-ionization-state metals \citep{Jaskot14,Henry15,Verhamme15}.
These UV properties are consistent with a low \hi\ content, 
leading to the ionizing radiation escape, as was theoretically demonstrated 
by  \citet{Verhamme15}, and observationally confirmed 
by \citet{Verhamme17} and \citet{Chisholm17}.

The \lya\ line of hydrogen (1215.67\,\AA) is one of the primary tools for 
detecting high-$z$ galaxies 
\citep[e.g.][]{Ouchi09,Sobral15,Matthee15,Zitrin15,Oesch16,Bagley17}. 
It also is a powerful tool to study conditions in the interstellar 
medium (ISM), both at 
low and high redshift \citep[e.g.][]{Hayes13,Hayes14,Verhamme15,Guaita17}. 
\lya\ resonantly scatters off neutral hydrogen, which results in strong 
radiative transfer effects both in real and frequency space. 
The escape of \lya\ from galaxies is a complex, 
multi-parameter problem: \lya\ photons trapped by scattering  
are more susceptible to dust absorption. On the other hand, 
outflows, low dust contents, and low \hi\ column densities help their escape 
\citep[e.g.][]{Kunth98,Shapley03,Atek08,
Verhamme08,Scarlata09,Wofford13,Hayes14,Hayes15,Henry15,Rivera15}. 
Building on the early analytical works \citep{Adams72,Neufeld90}, 
numerical \lya\ radiation transfer models were needed to demonstrate the  
effects of the ISM conditions on the \lya\ spectral profiles.  
Monte Carlo codes computing the \lya\ transfer in simplified 
plane-parallel or spherical geometries proved to be useful for this task 
at a relatively low computational cost 
\citep{Ahn01,Verhamme06,Dijkstra06,
Barnes10,Dijkstra12,Laursen13,Duval14,Zheng14,
Behrens14,Verhamme15,Dijkstra16,Gronke15,Gronke16,Gronke16b}. 
Application of the models to real galaxies has 
successfully reproduced most of the observed \lya\ profile features
\citep[e.g.][]{Verhamme08,Dessauges10,
Noterdaeme12,Krogager13,Hashimoto15,Martin15,Yang16,Yang17}. 

The similarity of green peas to high-redshift LBGs and LAEs 
\citep{Cardamone09,Izotov11,Jaskot13,Nakajima14,
Schaerer16} 
make them important low-$z$ laboratories for 
studying the LyC and \lya\ escape mechanisms, essential for understanding the
formation and evolution of galaxies across the cosmic history. 
Various aspects of the \lya\ emission in green peas were addressed by  
\citet{Jaskot14}, \citet{Henry15}, \citet{Yang16}, \citet{Yang17},
\citet{Verhamme15}, and \citet{Verhamme17}.
Radiative transfer models were used by \citet{Yang16,Yang17} to interpret 
\lya\ profiles of fifty-five GPs. Their modelling generally achieved 
acceptable fits between the single-shell homogeneous models and the GPs. 
In this paper, we reinterpret the archival HST  
Cosmic Origins Spectrograph (COS)
observations of \lya\ from twelve green peas studied by \citet{Yang16}. 
Using an independent radiative transfer code \citep{Verhamme06}, we extend 
this study by including observational ISM constraints. 
If we apply no constraints,
we reproduce the \citet{Yang16} results and successfully fit 
the observed \lya\ profiles. However, we show that the ISM 
parameters characterizing the best-fitting models   
are in disagreement with those measured from independent UV and optical data. 
If we impose the observational constraints in the fitting process,  
the homogeneous shell models do not provide good fits to the data. 
We evaluate the mismatches and discuss the validity of the models for these 
peculiar spectra.   

We structure the paper as follows.
We describe the data in Sect.\,\ref{sec_data}  
and the \lya\ radiative transfer models in Sect.\,\ref{sec_rt}. 
We present the model fitting results both with and without the 
application of observational constraints in Sect.\,\ref{sec_results}. 
We discuss the differences between modelled and observed ISM parameters, 
their correlations, and model limitations in Sect.\,\ref{sec_discussion}.

\section{Data sample}
\label{sec_data}

\begin{table}[t!]
\begin{center}
\caption{Green pea sample.}
\begin{tabularx}{0.4\textwidth}{cccc} 
\toprule
\toprule
ID     & RA  &  DEC     & $z$\tablefootmark{(a)} \\
\midrule
GP\,0303 & 03 03 21.41 & $-$07 59 23.2  & 0.16489 \\
GP\,0816 & 08 15 52.00 & $+$21 56 23.6  & 0.14095 \\ 
GP\,0911 & 09 11 13.34 & $+$18 31 08.2  & 0.26224 \\
GP\,0926 & 09 26 00.44 & $+$44 27 36.5  & 0.18070 \\
GP\,1054 & 10 53 30.80 & $+$52 37 52.9  & 0.25265 \\
GP\,1133 & 11 33 03.80 & $+$65 13 41.4  & 0.24140 \\
GP\,1137 & 11 37 22.14 & $+$35 24 26.7  & 0.19440 \\
GP\,1219 & 12 19 03.98 & $+$15 26 08.5  & 0.19561 \\
GP\,1244 & 12 44 23.37 & $+$02 15 40.4  & 0.23942 \\
GP\,1249 & 12 48 34.63 & $+$12 34 02.9  & 0.26340 \\
GP\,1424 & 14 24 05.72 & $+$42 16 46.3  & 0.18480 \\ 
GP\,1458 & 14 57 35.13 & $+$22 32 01.8  & 0.14859 \\
\bottomrule
\end{tabularx}
\\
\tablefoot{
$(a)$ Derived from Gaussian fitting of multiple SDSS 
emission lines. 
Adopted from \citet{Henry15}, and derived in a consistent way 
here for GP\,0816 and GP\,1458 that were not part of their sample.
We assume a conservative error of 40\,\kms\ due to wavelength calibration. 
}
\label{tab_sample}
\end{center}
\end{table}

\subsection{Observations and archival data}
\label{sec_obs}

\paragraph{\bf HST \lya\ data:}
We use archival far-ultraviolet (FUV) spectra of 
twelve green pea galaxies at redshift $z\!\!\sim\!\!0.2$ 
(Table~\ref{tab_sample}), 
observed with the Hubble Space Telescope (HST) 
under programmes GO\,12928 (PI A.~Henry), GO\,13293 (PI A.~Jaskot), 
and GO\,11727 (PI T.~Heckman).  
The \lya\ spectra were obtained with the $2.5\arcsec$ 
Primary Science Aperture (PSA) 
of the 
Cosmic Origins Spectrograph (COS) onboard the HST,
with the use of the medium resolution grism G160M. 
We use the standard pipeline-reduced data obtained through the 
Mikulski Archive for Space Telescopes (MAST). 
Details of the observations and various target properties
were included in \citet{Henry15}, \citet{Jaskot14}, and \citet{Heckman11}. 
Additional information on the target GP\,0926 from the HST imaging 
and ground-based spectroscopy are available in 
\citet{Basu-Zych09}, \citet{Goncalves10}, \citet{Hayes13}, 
\citet{Hayes14}, \citet{Rivera15},
and \citet{Herenz16}. 

The COS spectral resolution depends on the size of the source:
the resolving power varies between $R\!=\!16\,000$ ($<20$\,\kms) 
for a point source and $R\!=\!1500$ ($200$\,\kms) 
for a uniformly filled aperture (see the COS handbook). 
The COS acquisition near-UV images reveal that the green pea 
diameters are $\lesssim1\arcsec$ \citep{Henry15}. 
We assume that the \lya\ emission extent is typically a factor of 
two to\,four\,larger than the stellar continuum extent, based on the analysis 
of the GP cross-dispersion sizes in the two-dimensional COS spectra 
\citep{Yang17sizes}. 
A similar result was achieved by deep HST Advanced Camera for Surveys (ACS)
imaging of nearby star-forming galaxies \citep{Hayes13,Hayes14}. 
We estimate the COS resolution for GPs to be $\sim\!100$\,\kms,
under the assumption that the \lya\ emission distribution  
is peaked, as is the usual case in the known 
star-forming galaxies \citep{Hayes14}.
This is consistent with the \lya\ spectra appearance:  
sharp peaks and troughs on the one hand, and the lack of more detailed \lya\ 
sub-features on the other.
We have rebinned the COS spectra to a sampling of 25\,\kms.

\paragraph{\bf Ancillary HST UV measurements:}

Aside from \lya, the COS far-UV (FUV) spectra include 
a series of absorption lines of low-ionization-state (LIS) metals, 
such as \si. Due to the low ionization potential of these species, 
the lines 
provide valuable information on geometry and kinematics of the 
\hi\ gas, where \lya\ propagates. The LIS lines of our sample were analysed 
in \citet{Jaskot14}, \citet{Henry15}, and \citet{Yang16},
and we adopt here 
their kinematic parameters to describe the \hi\ medium for \lya\ modelling 
(Sect.~\ref{sec_constrained}).

\paragraph{\bf Ancillary optical SDSS data:}
We use the optical Sloan Digital Sky Survey (SDSS) spectra
to measure the emission-line redshifts 
\citep{Henry15}
and to estimate   
parameters of the intrinsic \lya\ line, that is, the line as it would appear 
before radiation transfer.
If both \lya\ and the Balmer lines are 
produced by the same recombination process, 
their respective luminosities are tied together through scaling laws, 
set by atomic physics. 
While \lya\ undergoes a resonant
radiative transfer in neutral hydrogen, Balmer lines travel through
the medium unaltered by \hi, and are only attenuated by dust. 
The observed \ha\ and/or \hb\ line profiles
corrected for the instrumental dispersion 
thus carry 
information about the intrinsic \lya\ line width. 
Their flux corrected for dust absorption then  
provides an estimate of the total produced \lya\ before it 
undergoes the radiative transfer. 
The SDSS spectra have spectral resolution 
$R\!\sim\!2000$ (150\,\kms), and were
obtained with circular $3\arcsec$ optical fibres, 
similar to the $2.5\arcsec$ COS aperture. 
The GPs are compact, unresolved in the SDSS, and therefore the 
difference between the COS and SDSS apertures is not significant 
(but see Sects.\,\ref{sec_rt} and \ref{sec_results} where we discuss 
the possible impacts).  
With the medium SDSS spectral resolution, the information about the 
intrinsic \lya\ is limited to the total flux and basic kinematics, of which 
we make full use in this paper. All of our data and models have been 
corrected for the instrumental dispersion.  
We describe the application of the constraints in more detail 
in Sect.\,\ref{sec_constrained}.

\subsection{Description of the sample: \lya\ line profiles}
\label{sect_profiles}

\begin{table}[b]
\caption{\lya\ spectral shape parameters.}
\label{tab_ewbr}
\small
\begin{tabularx}{0.46\textwidth}{lcccr}
\toprule
\toprule
ID       & $v_B$        & $v_R$  & Blue/Red                      & trough\hspace*{0.1cm} \\
         & [\kms]       & [\kms] & EW ratio & [\kms] \\
         & (1)          & (2)    & (3)                           & (4) \\
\midrule
GP\,0303 & $-300\pm60$  & $150\pm40$ &  0.06 & $-110\pm50$  \\ 
GP\,0816 & $-210\pm40$  & $140\pm50$ &  0.36 & $-20\pm40$   \\
GP\,0911 & $-290\pm50$  & $ 80\pm40$ &  0.17 & $-60\pm40$   \\
GP\,0926 & $-160\pm100$ & $220\pm120$&  0.14 & $-40\pm40$   \\
GP\,1054 & $-220\pm100$ & $200\pm40$ &  0.11 & $70\pm80$    \\
GP\,1133 & $-90\pm40$   & $230\pm60$ &  0.51 & $70\pm60$    \\
GP\,1137 & $-250\pm100$ & $170\pm50$ &  0.12 & $-20\pm40$   \\
GP\,1219 & $-80\pm40$   & $170\pm50$ &  0.38 & $20\pm40$    \\
GP\,1244 & $-240\pm40$  & $250\pm40$ &  0.33 & $-10\pm40$   \\
GP\,1249 & ---          & $80\pm40$  &  0.00 & ---          \\
GP\,1424 & $-150\pm60$  & $220\pm40$ &  0.67 & $70\pm40$    \\
GP\,1458 & $-360\pm60$  & $390\pm100$&  0.40 & $40\pm60$    \\
\bottomrule
\end{tabularx}
\tablefoot{
(1) Position of blue \lya\ peak measured from the systemic redshift; 
(2) Position of red \lya\ peak measured from the systemic redshift; 
(3) Blue-to-red equivalent width ration, measured with respect to the central
trough. Conservative 20\%\ error considered, due to 
continuum calibration; 
(4)  Position of central \lya\ trough measured from the systemic redshift.
}
\end{table}

We describe here the main features of the
observed \lya\ line profiles,
and derive the first implications for radiative transfer, 
independent of any model geometry.   

\paragraph{\bf Double-peaked emission lines:}
As already noted by \citet{Henry15} and \citet{Yang16}, all 
of the twelve green peas show \lya\ in net emission, 
which is unusual for other local
star-forming galaxy samples of
similar size \citep{Wofford13,Rivera15}. 
The net \lya\ emission is only prevalent in galaxy samples 
selected by their high FUV luminosity, 
the Lyman-break analogues \citep[LBAs, see][]{Heckman11,Alexandroff15}.  
Strangely, none of the GP \lya\ spectra have a 
P-Cygni profile with a redshifted emission and a 
blueshifted absorption,
which is often considered as the typical \lya\ line signature, mainly in 
high $z$.  
The \lya\ line is double-peaked in eleven of the twelve 
targets of our GP sample, and similar statistics are present in other 
GP samples, 
such as those of \citet{Verhamme17} and \citet{Yang17}, 
which is truly unusual for any other galaxy sample.
In high redshift, multiple-peak \lya\ profiles have drawn observers' 
attention only recently, after their discovery in low-$z$ galaxies, such as 
in \citet{Heckman11}, \citet{Martin15}, and \citet{Alexandroff15}.
The \lya\ line identification in distant LAEs was traditionally done
by its asymmetric single peak. 
\citet{Kulas12} and \citet{Trainor15} estimated the incidence 
of multiple peaks to be $\sim\!30$\% 
among UV-selected star-forming galaxies $z=2-3$ showing \lya\ emission.
The actual number can be higher, due to the blue peak attenuation 
by the inter-galactic medium (IGM) \citep{Laursen11,Dijkstra14}. 
Also, the typical spectral resolution in 
high-$z$ observations is lower than that of the HST/COS, and therefore 
some of the spectral profiles can in reality be double peaks, such as 
those with non-zero 
flux blueward of the systemic redshift \citep[][]{Erb14},

The relative equivalent width ($EW$) of the blue peaks varies across 
our sample, and it represents 5\,--\,65\% of the red peak $EW$
(Table~\ref{tab_ewbr}).  
To measure the blue and red $EW$s, we separated the \lya\ profile 
into two parts, divided by the central trough between the peaks.   
This differs from the definition previously used in the literature
\citep{Heckman11,Erb14,Henry15}, 
where separation between the blue and red parts of the spectra was defined
by the systemic redshift. 
Our definition reflects the sufficient data resolution and the need to 
characterize the individual peaks, independently of the 
redshift (see Discussion). 

We infer, from the systematically stronger red peak, that the GP \lya\ 
is transferred in outflowing media.  
This is consistent with the 
measurements of UV absorption lines that originate from
low-ionization state (LIS) metal species, which are 
blueshifted with respect to the systemic redshift, and thus 
indicate outflows. The red \lya\ dominance is 
model independent and has been illustrated analytically and numerically 
in various  
geometries such as slabs \citep{Neufeld90}, spherical shells 
\citep[e.g.][]{Verhamme06,Dijkstra06},  
and in radiative transfer coupled to  
full hydrodynamic simulations \citep[e.g.][]{Verhamme12}.
The double-peaked profiles may have various origins, 
such as \lya\ transfer in low-velocity media, in clumpy media or 
other, less studied, geometries. We will discuss this question 
in Sect.~\ref{sec_discussion}.

\paragraph{\bf \lya\ profile symmetries:}
We have (re-)measured the positions of the red and blue \lya\ peaks, and of the 
troughs that separate them (Table~\ref{tab_ewbr}). 
Without any modelling at this stage, we determined the local 
flux maxima and minima. 
 Uncertainties resulting from the peak shape and flux variations
were included in the error bars, together with the 
wavelength calibration uncertainties. 
In the cases where the \lya\ peak or trough had a multi-component character, 
or where the peak top had a peculiar shape with a varying flux 
(such as flat-top blue peaks with flux variations in GP\,1054 or GP\,1137),  
we computed the mean position of the components, and included their 
variance in the uncertainty.  
This definition, which was the best choice for the comparison with models,  
  may differ from that applied in 
\citet{Henry15} or \citet{Yang16}, and we find differences in some of the 
measurements. 
Nevertheless, the measurements are consistent 
with the independent papers within the stated error bars.

We provide the peak and trough positions in the form of velocity offsets,  
measured from the systemic redshift that was derived from the SDSS emission 
lines (Table~\ref{tab_sample}).  
We considered the combined wavelength calibration error of the SDSS 
and the HST/COS to be 40\,\kms\ \citep{Henry15}.
We find that the central \lya\ trough is redshifted with respect to the 
systemic redshift in five targets of the sample 
(GP\,1054, GP\,1133, GP\,1219, GP\,1424, and GP\,1457), that is usually those 
with strong blue peaks (Table~\ref{tab_ewbr}). 
In at least two of them (GP\,1133 and GP\,1424) the shift cannot be explained 
by measurement errors, 
which include the systematic uncertainty on the wavelength calibration, 
noise on the spectral line profile, and additional features in the trough.
These two targets also have the 
largest relative flux of the blue peak. The redshifted \lya\ troughs 
are surprising and were not 
expected in galaxies with \hi\ outflows: due to the Doppler shift in the 
outflows, the largest \lya\ optical depth should be shifted from zero to   
negative velocities \citep{Verhamme06}. 
In a reversed configuration, an inflow, the trough would be placed in 
positive velocities. However, the blue peak should dominate 
the flux in that case, which is not what we observe in the GPs. 
The inflow scenario therefore seems improbable.

\begin{figure}
\begin{tabular}{l}
\includegraphics[width=0.45\textwidth]{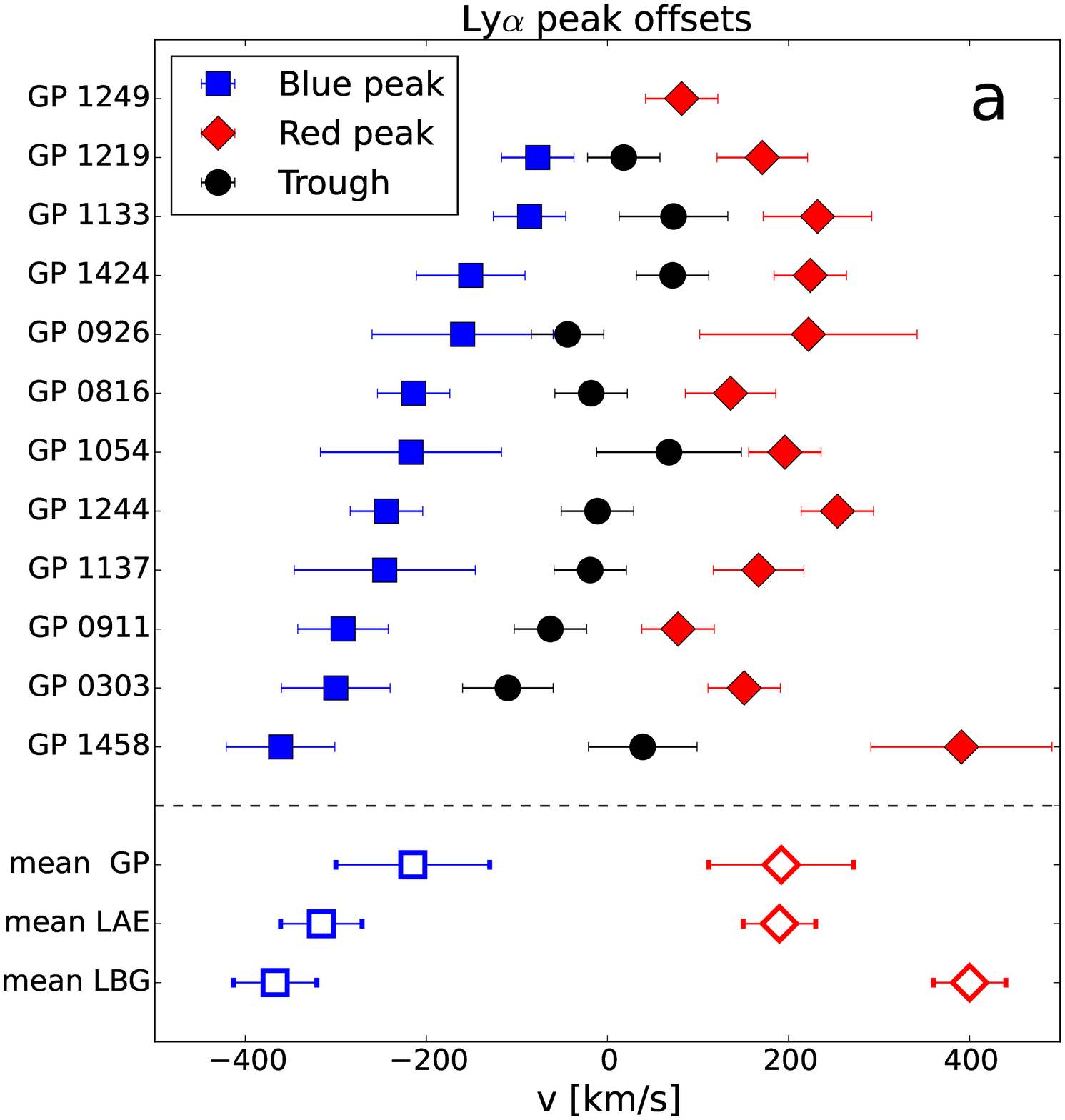}
\\
\includegraphics[width=0.45\textwidth]{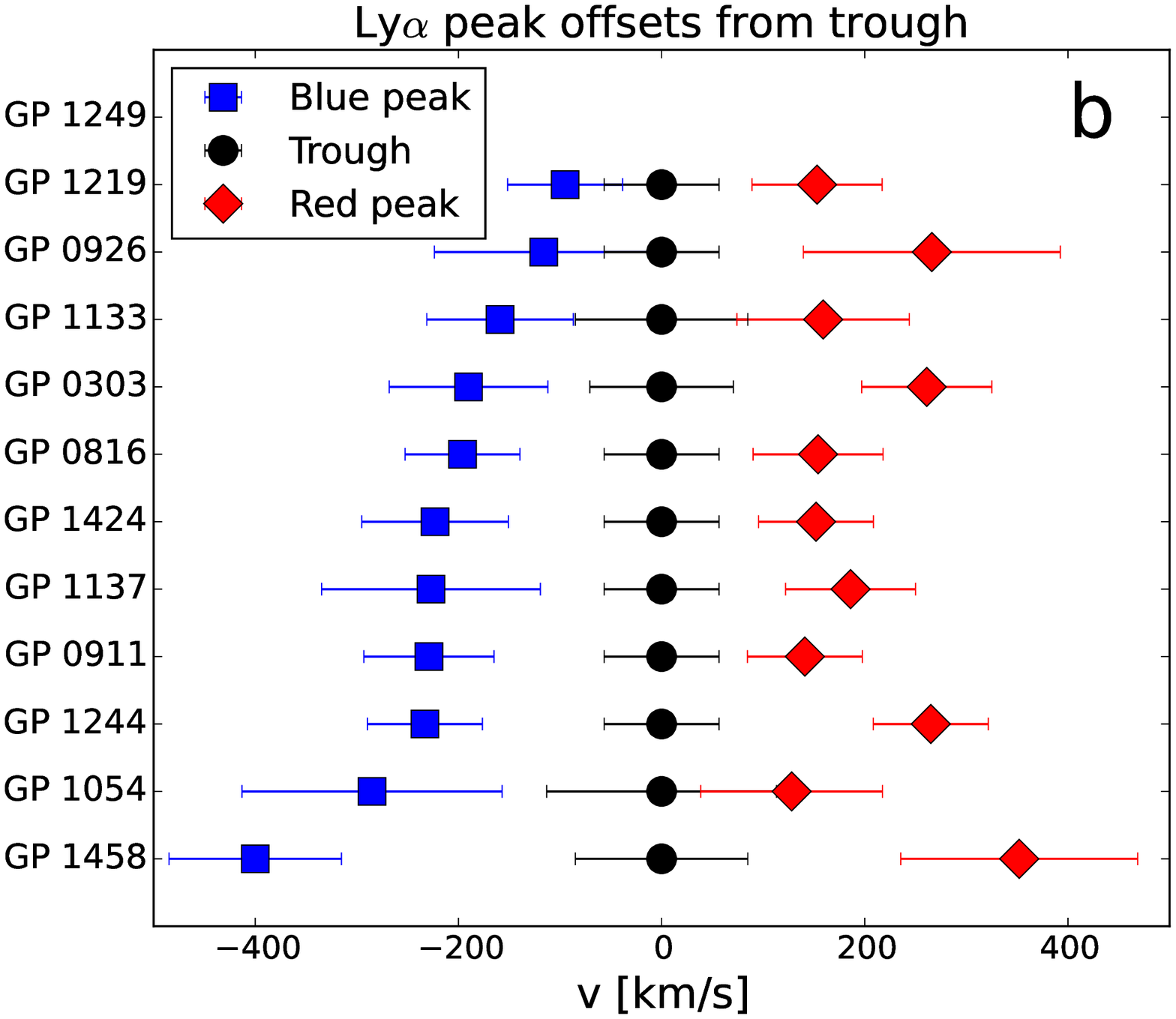}
\\
\end{tabular}
\caption{Asymmetry of the blue and red \lya\ peaks in the GP sample, measured
from a) the systemic redshift, and b) the central trough. 
Panel (a) also includes the mean
positions of the blue and red peaks for 
this GP sample, and for LAE and LBG samples drawn from the literature 
\citep{Hashimoto15}. The GPs are similar to the LAEs in the red peak 
positions, but have a smaller mean blue offset. 
}
\label{fig_symm}
\end{figure}

We illustrate the peak and trough positions in Fig.\,\ref{fig_symm}, 
where the targets have been sorted by their blue peak offset, measured
either with respect to the systemic redshift (Fig.\,\ref{fig_symm}a) or
to the central trough position (Fig.\,\ref{fig_symm}b). 
The red and blue peak positions are not symmetric. 
Ordering of the targets by their blue peak offset 
did not produce any alignment in the red peak offset.  
We will see in Sect.\,\ref{sec_sym_obsmod} that the peaks are not equally 
broad either, and we will discuss the implications 
for the radiative transfer models.

\paragraph{\bf Low column density of neutral hydrogen:}
The GP \lya\ profiles are unusually narrow, with a
small separation between their peaks 
($400\pm100$\,\kms). We show in Fig.\,\ref{fig_symm}a that this separation
is smaller than in typical high-$z$ LAEs ($510\pm60$\,\kms) 
and LBGs ($770\pm60$\,\kms), drawn from the \citet{Hashimoto15} sample.  
While the major difference between LBG and LAE double-peaked profiles is in the 
red peak position ($400\pm40$\,\kms\ in LBGs, $190\pm40$\,\kms\ in LAEs), 
it is the blue peak that drives the difference between the GPs 
and the high-$z$ samples
($-370\pm50$\,\kms\ in LBGs, $-320\pm50$\,\kms\ in LAEs, and $-210\pm90$\,\kms\ in GPs). 
The observed small separations show that \lya\
is able to escape close to the systemic redshift and that the effects
of radiative transfer are relatively weak, irrespective of the 
model geometry \citep{Verhamme15}.
Several ISM parameters can contribute to achieve this condition: 
low \nh, large \hi\ velocities, or a clumpy medium with low-density 
inter-clump gas. 
Observations have proved that the \lya\ peak offsets are small in 
galaxies with escaping Lyman continuum 
\citep{Izotov16,Izotov16b, Izotov18} and that the double-peak separation 
correlates with the LyC escape
fraction \citep{Verhamme17}. The LyC escape requires a low \hi\ column 
density along the line of sight, \nh\,$\lesssim10^{18}$\,cm$^{-2}$ if the
ISM is homogeneous. 
The \lya\ spectra of our GP sample are similar to those of the 
LyC leakers, therefore 
there is a high probability that the GPs studied here have analogously 
low \hi\ column densities (no direct confirmation exists, no LyC observations 
are available for the present sample).

A large opacity in the \lya\ line core produces 
a trough separating the red and blue 
peaks. However, the trough does not reach the zero flux level  
in approximately half 
of our sample (GP\,0816, GP\,0911, GP\,1133, GP\,1219, and GP\,1424), 
confirming that the opacity (and thus \nh) is surprisingly low. 
\citet{Yang16} found a correlation between the residual flux in the 
trough and the \lya\ escape fraction, which suggests   
that the non-zero flux is not an artificial effect caused by an insufficient
spectral resolution. 
Sub-features can be seen in the central trough of GP\,1133: two minima
separated by 80\,\kms. 
This could indicate that the line core is either partially 
refilled with emission, or that the blue minimum 
corresponds to the deuterium absorption, as discussed in 
\citet{Dijkstra06} and \citet{Verhamme06}.

Another confirmation of the low \nh\ in the GPs comes from 
the absence of underlying broad \lya\ absorption,
analogously to $z\!\sim\!0.3$ LyC leakers \citep{Verhamme17}.
Star-forming galaxies with observed \lya\ emission lines commonly show 
an underlying absorption trough, with wings visible on a much wider 
wavelength scale than the emission spectrum 
\citep[see e.g.][]{Wofford13,Rivera15,Duval16,Schaerer08,Dessauges10,Quider10}. 
From the modelling point of view, as \lya\ resonantly scatters off \hi, 
the absorption part of the spectrum is the result of the  
photon removal from the line of sight, while emission is produced by 
photons scattered out of resonance.  
The absorption trough becomes deeper and broader with the 
increasing column density of the foreground \hi, and with aperture 
losses \citep{Verhamme06,Verhamme17}. 
We have done a careful inspection of the continuum in all twelve GPs, and found no signs 
of absorption, except for a shallow trough in GP\,1458 and 
possibly in GP\,0303. The absence points to generally low \hi\ column 
densities in GPs, 
likely similar to those in LyC leakers of \citet{Izotov16,Izotov16b,Izotov18}.

\paragraph{\bf Comparison between \lya\ and \hb\ profiles:}
If produced by pure recombination, the intrinsic \lya\ 
line profile (before radiative transfer) should share the kinematic 
characteristics of the Balmer lines, and be their scaled version. 
Radiative transfer effects transform the \lya\ profile by removing photons
most strongly from the line centre, redistributing them to the wings, or 
destroying them by absorption. 
The resultant profile is broadened, 
attenuated, and with shifted peaks.  
Figures \ref{fig_constrained} and \ref{fig_fits} show a 
direct comparison between the observed \lya\ profile and 
\hb\ scaled by the Case~B factor of 23.5 \citep{Dopita03}. 
The figures also present the results of model fitting  
(Sect.\,\ref{sec_results}), while here we 
focus on the comparison of the observational profiles. 
We used \hb\ instead of \ha,\ which partially blends with the \nii\ lines and 
would impede studying the line wings. 
We present the data with subtracted continuum, 
which was fit by a first order polynomial in the vicinity of \hb\ and \lya.
We corrected both \hb\ and \lya\ fluxes for the Milky Way (MW) 
extinction \citep{Schlafly11} using the \citet{Cardelli89} extinction law.
The MW extinction values were obtained from the NASA Extragalactic Database
(NED) and were listed in \citet{Henry15} and \citet{Yang16}.
The \hb\ flux was additionally corrected for internal extinction
using the SDSS \ha\ and \hb\ fluxes with the assumption of their intrinsic 
ratio of 2.86 \citep{Dopita03}, and 
using the \citet{Cardelli89} extinction law. 
The corrected \hb\ flux was then used to approximate
the intrinsic \lya\ line flux. 
No correction for internal extinction was applied to \lya,
as it is not attenuated in the same way as the optically thin 
lines, and its radiative transfer includes effects of both gas and dust. 

The observed \lya\ blue wings are as broad 
as those of the scaled \hb\ line in most of the sample. 
For two targets, GP\,0303 and GP\,0911, such a similarity is also seen in 
the red wing. This is unusual, and signifies that the \lya\ profile has not 
been much broadened by radiative transfer.      
Remarkably,  \citet{Martin15} draw a similar conclusion for a sample of 
low-$z$ 
ultraluminous infrared galaxies (ULIRGs). 
We note that SDSS resolution is worse than that of COS. However, we have tested
that degrading the COS spectra to the SDSS resolution does not change our 
conclusion about the wings.

\begin{table*}[ht]
\begin{center}
\caption{
\label{tab_params}
Measured and model-fit parameters.
}
\tiny
\def\arraystretch{1.5}
\begin{tabularx}{1.0\textwidth}{c|cccc|cccccccc} 
\toprule
\toprule
&
\multicolumn{4}{c}{Observed ISM parameters (HST/COS, SDSS)} & 
\multicolumn{6}{|c}{Shell model results from unconstrained fitting}  
\\
\midrule
ID 
    
    & $v_\mathrm{LIS}$ 
    & FWHM(H$\beta$) 
    & $\tau_\mathrm{d,obs}$
    & EW$_0$(\lya)$_\mathrm{obs}$ \ 
    & $\Delta z$
    & \vexp\  
    & $b$
    & $\log N_\mathrm{HI}$
    & $\tau_\mathrm{d}$
    & FWHM$_0$(\lya)
    & EW$_0$(\lya) 
\\
    & [\kms] 
    & [\kms] 
    & FUV 
    & [\AA]
    & [\kms]
    & [\kms]
    & [\kms]
    & [cm$^{-2}$]
    & FUV
    & [\kms]
    & [\AA]
\\  
 (1)      & (2)         & (3) & (4) & (5)       & (6)      & 
(7)       & (8)         & (9)           & (10)    &(11)  & (12)   \\ 
\midrule
GP\,0303  
& $-200^{+80}_{-80}$   & 160 &  0         & 150$^{+50}_{-50}$ &
$60^{+60}_{-0}$ & $150^{+50}_{-0}$ &$20^{+30}_{-10}$  
& $19.0^{+0.3}_{-1.2}$  &$1^{+2}_{-1}$  
&$500^{+100}_{-120}$ &$15^{+5}_{-5}$
\\
GP\,0816
& $-300^{+100}_{-100}$      & 130 &  0.4$^{+0.2}_{-0.2}$  & 200$^{+100}_{-100}$ &
$10^{+50}_{-0}$  & $20^{+30}_{-0}$ & $20^{+0}_{-10}$
& $19.0^{+0.3}_{-1}$ &$0.2^{+0.8}_{-0.2}$  
&$400^{+100}_{-0}$   &$75^{+25}_{-25}$
\\
GP\,0911 
& $-250^{+50}_{-50}$      & 220 &  2.6$^{+1.4}_{-1.4}$    & 140$^{+110}_{-110}$  &
$30^{+0}_{-70}$ & $100^{+50}_{-0}$ & $20^{+20}_{-0}$
& $18.5^{+0.5}_{-0.7}$   &$1.5^{+1.5}_{-0.5}$ 
&$500^{+0}_{-100}$ &$50^{+25}_{-0}$
\\
GP\,0926  
& $-280^{+100}_{-100}$          & 170 &  1.5$^{+0.8}_{-0.8}$   & 120$^{+70}_{-70}$  
&$140^{+5}_{-50}$& $150^{+50}_{-0}$   & $20^{+0}_{-10}$
& $18.5^{+0.5}_{-0.7}$ &  $1^{+1}_{-1}$   
&$550^{+50}_{-50}$  & $40^{+10}_{-10}$
\\ 
GP\,1054 
& $-180^{+80}_{-80}$       & 230 &  1.3$^{0.7}_{-0.7}$   & 140$^{+50}_{-50}$  
&$150^{+40}_{-40}$ & $150^{+0}_{-0}$   & $40^{+40}_{-0}$
& $19.3^{+0}_{-1.5}$   &$1^{+2}_{-1}$    
&$150^{+250}_{-50}$  &$20^{+10}_{-10}$
\\
GP\,1133 
&  ---          & 160 &  0.2$^{+0.1}_{-0.1}$   & 100$^{+50}_{-50}$  
&$160^{+50}_{-0}$ & $50^{+0}_{-30}$ & $10^{+0}_{-0}$
& $18.5^{+0.5}_{-0}$  &$0.5^{+2.5}_{-0.5}$ 
&$600^{+0}_{-100}$  &$35^{+5}_{-5}$ 
\\
GP\,1137 
& $-150^{+50}_{-50}$       & 210 &  1.0$^{+0.6}_{-0.6}$  & 200$^{+100}_{-100}$ &
$100^{+0}_{-50}$  & $150^{+100}_{-0}$ & $40^{+120}_{-20}$
& $18.7^{+0.3}_{-2.7}$    &$1^{+0.5}_{-1}$ 
&$550^{+60}_{-300}$ &$40^{+10}_{-10}$
\\ 
GP\,1219 
&  ---         & 160 &  0.03$^{+0.02}_{-0.02}$ & 200$^{+100}_{-100}$ 
&$115^{+50}_{-0}$ & $50^{+50}_{-0}$  & $20^{+0}_{-10}$
& $18.0^{+0.5}_{-2}$  &$0.2^{+0.8}_{-0.2}$   
&$600^{+0}_{-100}$ &$125^{+0}_{-25}$ 
\\
GP\,1244 
& $-90^{+50}_{-50}$      & 170 &  1.1$^{+0.6}_{-0.6}$  & 400$^{+300}_{-300}$ 
&$150^{+0}_{-100}$  & $150^{+0}_{-100}$ & $40^{+0}_{+0}$
& $18.5^{+1.1}_{-0.5}$ &$0.5^{+1.6}_{-0.3}$   &$1000^{+0}_{-300}$ &$75^{+25}_{-25}$
\\
GP\,1249 
& $-180^{+50}_{-50}$      & 160 &  0.7$^{+0.2}_{-0.2}$  & 200$^{+100}_{-100}$  
&$80^{+50}_{-50}$ & $250^{+0}_{-50}$ & $10^{+10}_{-0}$
& $16.0^{+2}_{-0}$   &$2^{+2}_{-1}$     
&$400^{+0}_{-0}$  &$75^{+5}_{-0}$
\\
GP\,1424 
& $-240^{+50}_{-50}$         & 210 &  0.7$^{+0.4}_{-0.4}$  & 200$^{+100}_{-100}$ 
&$120^{+50}_{-0}$ & $50^{+50}_{-30}$ & $20^{+20}_{-10}$
& $19.0^{+0.3}_{-1.5}$   &$0.2^{+0.8}_{-0.2}$ 
&$700^{+30}_{-100}$ &$75^{+40}_{-25}$ 
\\
GP\,1458 
&  $-30^{+70}_{-70}$          & 130 &  0.4$^{+0.3}_{-0.3}$ & 400$^{+300}_{-300}$
&$55^{+0}_{-45}$ & $20^{+0}_{-0}$ & $40^{+0}_{-20}$
& $20.2^{+0.3}_{-0}$ &$1^{+0.5}_{-0.5}$   
&$400^{+100}_{-250}$  &$100^{+200}_{-50}$\\
\bottomrule
\end{tabularx}
\tablefoot{
(1) Identifiers; 
(2) Mean LIS line velocities with their 1$\sigma$ error, 
adopted from \citet{Henry15}, and measured here for GP\,0816 
and GP\,1458;
(3) $FWHM$ of narrow \hb\ component, 
determined by double-Gaussian fits to SDSS line profiles, 
and corrected for instrumental dispersion.
We estimate a fitting error of 50\,\kms;
(4) FUV optical depth of dust 
near the wavelength of \lya, derived from \hb\ flux,
FUV continuum flux, and ISM extinction; 
(5) Intrinsic \lya\ equivalent width derived from \hb\ flux,
FUV continuum flux, and ISM extinction;     
(6) -- (12) Best-fitting model parameters, obtained from unconstrained \lya\ fitting. 
Median, 10$^\mathrm{th}$ percentile and 90$^\mathrm{th}$ percentile values of 20 models with the lowest $\chi^2$ for each target; 
(6) $\Delta z = (z_{\mathrm{fitLy}\alpha} - z_\mathrm{SDSS}) c$, where
 $z_{\mathrm{fitLy}\alpha}$ is the redshift of the best unconstrained 
 \lya\ fit, 
 $z_\mathrm{SDSS}$ is the redshift derived from SDSS emission line fits 
 (Table\,\ref{tab_sample}), and $c$ is the speed of light; 
(7) Shell expansion speed; 
(8) Doppler parameter; 
(9)  Neutral hydrogen column density; 
(10) FUV optical depth of dust 
in the vicinity of \lya; 
(11) $FWHM$ of intrinsic \lya\ line; 
(12) Equivalent width of intrinsic \lya\ line. 
}
\end{center}
\end{table*}

\section{Radiative transfer models}
\label{sec_rt}

\subsection{Code and model grid}

We use an enhanced version of the {\tt MCLya} 3D Monte
Carlo code of \cite{Verhamme06}, which
computes radiative transfer of the \lya\ line and the adjacent
UV continuum for an arbitrary three-dimensional (3D) geometry and velocity field.
We here assume the geometry of an expanding, homogeneous,
spherical shell composed of neutral hydrogen and dust, uniformly mixed.
A starburst region, which is the source of the \lya\ and UV continuum photons,  
is placed at the centre of the sphere filled with ionized gas
and surrounded by the neutral shell. The photons are 
collected after their propagation through the neutral shell in all directions.  
The expanding geometry was motivated by the \hi\ outflows, which we detect in the
studied galaxies \citep{Henry15} and which additionally seem to be 
ubiquitous at both low- and high-redshifts 
\citep[e.g.][]{Shapley03,Wofford13,Rivera15,Chisholm15}.
The spherical configuration was motivated by the observation of 
superbubbles in star-forming galaxies, as 
described by for example \citet{Tenorio99} and \citet{Mas-Hesse03}.
The {\tt MCLya} models have previously been successfully applied to 
reproduce the \lya\ spectra of low- and high-$z$ galaxies 
\citep{Verhamme08,Schaerer08,Dessauges10,
Vanzella10,Lidman12,Leitherer13,Hashimoto15,Duval16,Patricio16}.

The modelled shell is characterized by four parameters: the radial expansion
velocity \vexp, the \hi\ column density \nh, 
the optical depth $\tau_\mathrm{d}$ of dust at wavelengths in the vicinity 
of \lya, 
and the \hi\  Doppler parameter $b$ that describes the internal 
shell kinematics including thermal and turbulent velocities.
Photons from the source have an initial 
frequency distribution, and the path and the frequency change through the shell 
are followed for each of them.      
A grid of $>\!6000$ synthetic models was constructed by varying the 
shell parameters and running the full {\tt MCLya} simulation for each parameter
set \citep{Schaerer11}. The resulting spectra were then  
post-processed to account for the intrinsic line profile, and were further 
smeared to match the observed spectral resolution estimated from the \lya\ 
source extent.

We define the initial, intrinsic \lya\ profile assuming  
a flat stellar UV continuum and a \lya\ emission line, which 
can either be a Gaussian, a double-Gaussian, or another function, such
as the observed Balmer line profile.   
If we assume a Gaussian profile, the model 
adjustment has seven free parameters. Four parameters describe the \hi\ shell: 
\vexp, \nh, $b$, $\tau_\mathrm{d}$;  
and three parameters describe the Gaussian line: the line width \fwhmlyai,
the equivalent width \ewlyai, and the line centre, that is, the redshift $z$. 
Each of the fitting parameters regulates a different measurable 
characteristics of the 
resultant \lya\ profile \citep{Verhamme06,Gronke15}. 
The profiles are most sensitive to \vexp\ and \nh, 
both of which determine the offset of the resultant \lya\ emission peak from 
the restframe position, while \vexp\ also determines the $EW$  
of the \lya\ blue peak. 
The intrinsic \lya\ equivalent width 
\ewlyai\ and the dust optical depth $\tau_\mathrm{d}$ 
regulate the \lya\ flux scaling, and a large
dust content usually results in a prominent absorption trough 
in the blue part of the profile if the medium is expanding. 
Large \nh\ and $\tau_\mathrm{d}$ give rise to broad \lya\ absorption profiles. 
The role of $b$ is less straightforward to describe and is usually degenerate 
with other parameters.  

We use an automated line profile fitting tool \citep{Schaerer11}, 
which explores the entire grid or its defined part. 
The fitting parameters can either be constrained to definite values or 
to intervals of values. 
The fitting procedure then searches for the minimum $\chi^2$ and 
computes the $\chi^2$ maps. 
We here define intervals for the fitting parameters that either correspond to 
the observational uncertainties (Sect.\,\ref{sec_constrained}) 
or to the entire grid extent (Sect.\,\ref{sec_unconstrained}). 
We keep twenty best-fitting models for each fitting run and each target
to have a better control of the fitting parameter stability. 
We describe their use for the different fitting approaches   
in the respective sections.  

\subsection{Constraining the free parameters}
\label{sec_constrpar}
Depending on the availability of ancillary data, several of the fitting 
parameters can be constrained. 
In the present sample, we were able constrain up to five parameters 
using the optical SDSS and UV HST/COS data (Tables \ref{tab_sample} and 
\ref{tab_params}):  
\begin{enumerate}
\item
Redshift $z$ (Table~\ref{tab_sample}):  
derived from Gaussian fits of $\sim\!20$ SDSS emission lines 
with a precision better than $10$\,\kms\ \citep{Henry15}. 
Applied to the \lya\ line, the redshift uncertainty is dominated by the 
wavelength calibration errors, $\sim\!40$\,\kms\ (Sect.\,\ref{sec_obs}).
\item
Expansion speed \vexp\ of the shell: measured from the UV absorption line
offsets of low-ionization species, 
which are commonly observed blueshifted, that is, outflowing. 
We report 
the characteristic LIS line velocities \vlis\ in Table~\ref{tab_params}. 
The velocities measured by \citet{Henry15} show 
variations between different transitions, which can be attributed to either noise or different amounts of emission filling \citep{Scarlata15}.
We list here their mean and consider 
the full range of velocities reported in that paper 
in order to give our \lya\ models
the best chance to match the observations. We have  
measured the missing velocities.
\item
Optical depth of dust absorption in the FUV, $\tau_\mathrm{d}$: estimated
from the SDSS Balmer line ratios by extrapolation to the FUV 
domain using extinction laws from the literature.
An inherent uncertainty associated with this estimate arises from the
uncertainty in the extinction law -- the commonly used laws 
give similar predictions
in the optical range, but dramatically vary in the FUV. 
We report in Table~\ref{tab_params} the mean value and variance 
of $\tau_\mathrm{d}$ derived from the extinction laws of 
\citet{Cardelli89}, \citet{Calzetti94}, and \citet{Prevot84}.
\item
Full width at half maximum of the intrinsic \lya\ line, \fwhmlyai, 
is assumed to be identical to that of \hb\ 
if both lines are formed by the same recombination mechanism. 
We fitted the SDSS \hb\ line profiles with two-component Gaussian 
functions, narrow and broad. 
We list the width of the dominating narrow component, \fwhmhb, in 
Table~\ref{tab_params}, and the broad component width and flux contribution 
in Table~\ref{tab_2G}. The line widths have been corrected for the 
instrumental dispersion. We assume an uncertainty of 100\,\kms\ in the \fwhmhb.
\item
Intrinsic \lya\ equivalent width, \ewlyai: 
estimated from the observed \hb\ flux and the observed 
FUV continuum, assuming that both lines were formed by recombination. 
Pure Case B recombination at $T=10\,000$\,K 
and $n_\mathrm{e}=10^{-2}$\,cm$^{-3}$ predicts the \lya/\hb\ 
flux ratio of 23.5 \citep{Dopita03}. 
Both the measured line and continuum fluxes were corrected
for dust attenuation, and therefore the uncertainties in the extinction law 
induce uncertainties in \ewlyai. In addition, the differences between
the nebular and continuum attenuation 
\citep{Calzetti97,Price14},
the possible need for aperture
corrections accounting for the differences between SDSS and COS, and 
the possible deviations in the temperature gave rise to the range of \ewlyai\ 
for each target reported in Table~\ref{tab_params}. 
\end{enumerate}

No constraints are available for the \hi\ column \nh, 
and the Doppler parameter $b$. To constrain $b$, high-resolution absorption 
lines would be necessary, resolving the individual \hi\ clouds.   
Therefore, $b$ and \nh\ are considered as free fitting parameters, 
ranging across the entire grid of models: $10$\,\kms\,$\leq b \leq 160$\,\kms, 
and $10^{16}$\,cm$^{-2}$\,$\leq$\,\nh\,$\leq10^{23}$\,cm$^{-2}$. 

\bigskip

For those parameters that require a comparison between the SDSS 
and the HST/COS data, 
we need to consider the differences between the spectrographs: 
1) the aperture size, and 2) the spectral resolution.
The different aperture sizes,  
2.5\arcsec\ in the HST/COS versus 3\arcsec\ in the SDSS, potentially  
affect the measured fluxes, 
and can therefore impact the \ewlyai\ derived from \hb.    
However, the near-UV HST/COS acquisition images \citep{Henry15} 
show that the GPs
are extremely compact, smaller than the aperture size. 
The GP optical images are unresolved in the SDSS, but we expect 
a similar morphology and extent as in the near-UV  
\citep[based on][]{Hayes13,Hayes14}. 
Nevertheless, to account for the possibility that the true optical emission 
is much larger than the acquisition image and that the aperture size matters
for the derivation of the intrinsic \lya,  
we considered both situations, with and without the aperture correction
to derive \ewlyai\ from \ewhb. 
Both cases were reflected in the    
\ewlyai\ intervals that we list in Table~\ref{tab_params}.
No correction was applied to the observed \lya\ emission due to its 
unpredictable nature. It is 
possible that a part of \lya\ is transferred outside the COS aperture, which 
we will further discuss in the following sections.  
Concerning the spectral resolution, it plays a role 
in the derivation of the intrinsic \lya\ profile from \hb. 
The SDSS resolution is worse (150\,\kms) than the HST/COS resolution  
for \lya\ ($\sim\!100$\,\kms, Sect.\,\ref{sec_data}). 
With the given SDSS resolution, our
\hb\ application is restricted to  
the instrument-corrected $FWHM$, with no other details about the line profile. 
To account for the COS resolution in the fitting process, we adjust the   
\lya\ grid model spectra correspondingly. 
We show in the following sections that the observed \lya\ spectra would be 
better reproduced with an input line broader than the SDSS profile, which cannot be explained by the difference in spectral resolution. On the other hand, 
other inconsistencies between models and data could potentially be better 
understood with the aid of high-resolution spectroscopy.

\section{\lya\ line profile modelling}
\label{sec_results}

\begin{figure*}[t!]
\includegraphics[width=1.0\textwidth]{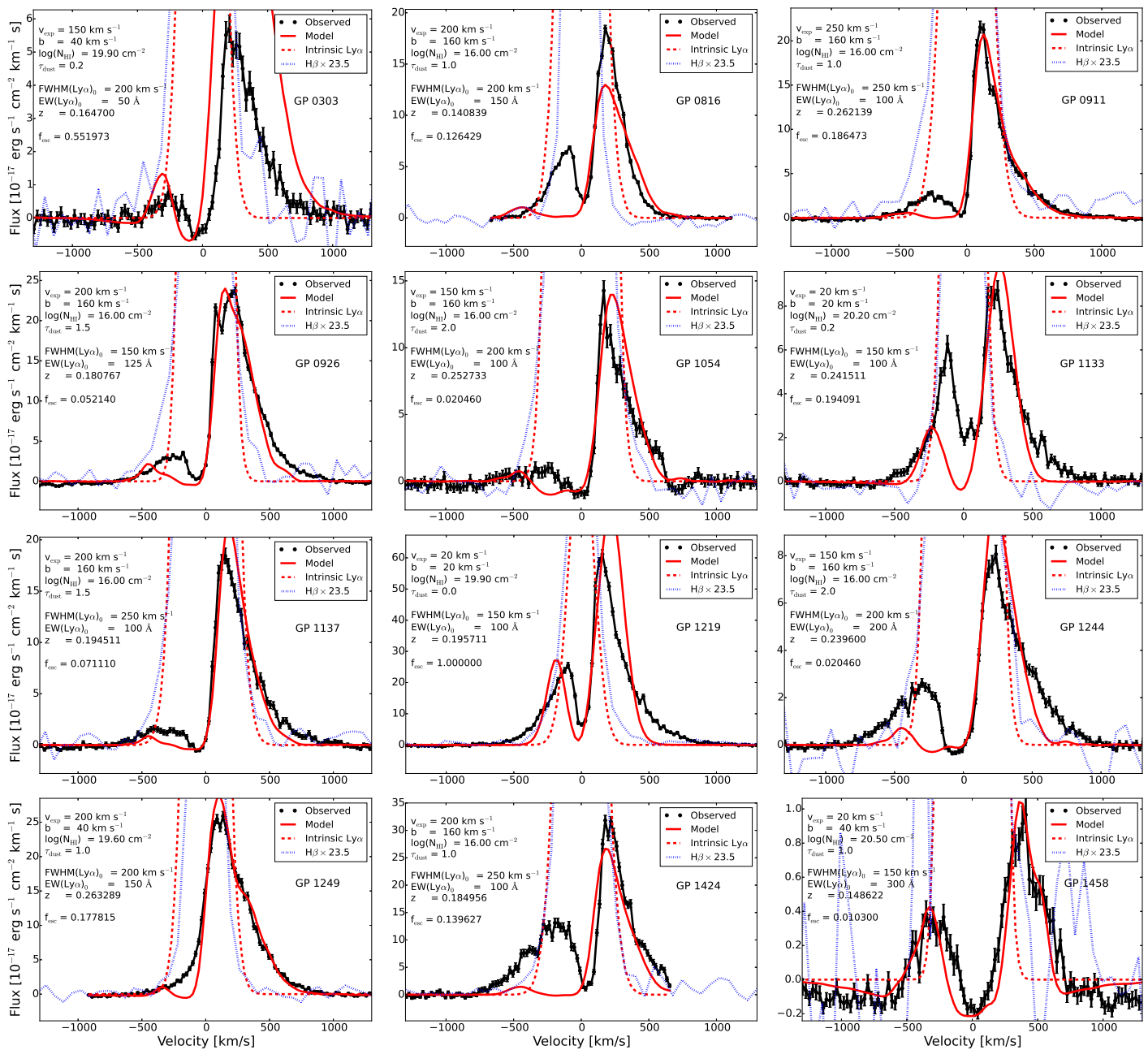}
\caption{Best-fitting \lya\ models with applied constraints from 
ancillary observational data (Sect. \ref{sec_constrained}). 
COS data and their errorbars are shown in black, 
the best-fitting model with the solid red line, the modelled 
intrinsic single-Gaussian \lya\  
profile with the dashed red line, and the observed SDSS \hb\ scaled by 
Case B factor 23.5 with the dotted blue line.
Parameters of the best-fitting model are shown in each panel. 
}
\label{fig_constrained}
\end{figure*}

We have carried out \lya\ profile modelling with the use of the model grid described 
in Sect.\,\ref{sec_rt} and with the application of
observational constraints. As we will see in Sections~\ref{sec_constrained} 
and \ref{sec_2G}, 
the constrained models do not reproduce all the features of the observed profiles, 
and we will therefore explore their failure and will relax all the constraints  
in Sect.~\ref{sec_unconstrained}.

\subsection{\lya\ profile fitting with constraints} 
\label{sec_constrained}

We have described the derivation of the constraining parameters
in Sect.\,\ref{sec_rt}.
Assuming a Gaussian profile of the intrinsic \lya\ line, with the $FWHM$ equal
to the dominant, narrow component of \hb, 
five of the seven fitting parameters can be constrained. 
For fitting purposes, the parameters were assigned the 
intervals listed in columns 2\,--\,5 of Table~\ref{tab_params}, which 
reflect the observed values and their uncertainties.  
Redshifts and line widths were 
considered with conservative errors of 50\,\kms\ and 100\,\kms, respectively, 
to account for 
the COS wavelength calibration and for the SDSS line decomposition into 
Gaussians.   
The \vexp\ velocity remains a  free fitting parameter in two targets, 
GP\,1133 and GP\,1219, where no UV LIS lines have been detected.  
The \lyb\ line is available, but its proper modelling would be 
necessary to disentangle the stellar and ISM contributions, therefore we have 
not used it.
Before the \lya\ fitting, we convolved the synthetic 
grid spectra with a Gaussian 
function, which simulated the finite spectral resolution, 100\,\kms\ 
(Sect.\,\ref{sec_data}).  
We carried out a series of tests to check how the spectral resolution 
would affect the fits. Model spectra convolved with too broad or too narrow 
profiles were not compatible with the observed 
features of the \lya\ profile (not sharp enough peaks and troughs, 
or too detailed substructure, respectively). A good match was achieved 
at $\sim\!100\pm30$\,\kms, which is consistent with
the COS estimation derived from the target morphology. 
We note that this convolution was missing in the work of 
\citet{Yang16}, which may explain some of the differences between their results
and this work.     
 
We have run the fitting tool on the model sub-grid defined by the constraints, 
keeping twenty models with the lowest $\chi^2$ for each target.  
The twenty models served as a visual check of the model matching.
We plot the data and the lowest-$\chi^2$ model 
for each galaxy in Fig.\,\ref{fig_constrained}. 
We overplot the theoretical 
intrinsic \lya\ line, derived from the best-fitting model, 
and the observed SDSS \hb\ line
scaled by the Case~B factor of 23.5. 
To enable the comparison, all observational and modelling data were 
converted to velocity units, and the respective 
continuum was subtracted from each line. The intrinsic
\lya\ model was convolved with a Gaussian profile corresponding to the SDSS 
resolution for a direct comparison with the SDSS data.
The best-fitting model parameters are listed in each panel. 
Any differences between the model parameters and the ancillary data 
parameters, or between the intrinsic \lya\ and \hb\ profiles, 
are due to the observational uncertainties and the discreteness of the model 
grid.

We immediately notice large discrepancies between the fits and the data. 
The models dramatically fail in reproducing the blue peak for almost every 
target, in both flux and position. 
The match for the red peaks is better and can be considered satisfactory in 
approximately half of the sample, while the peak positions and widths 
disagree with the observations in the remaining targets. 
Furthermore, the central trough positions of the double-peaked models are 
offset with respect to the observed troughs.    
The homogeneous shell models regulate the relative blue peak flux by \vexp, 
while the trough position is the location of the maximum optical depth, 
dependent on $z$ and \vexp, and cannot be forced by any other combination 
of the remaining parameters.
With $z$ and \vexp\ fixed by the observational constraints, 
the fitting algorithm can only attempt to reproduce the  
peak positions by varying \nh, which was a free parameter here, but cannot
produce the correct peak flux ratios and trough positions. 
In the logic of the homogeneous models, the ancillary data did not provide the 
correct $z$ and \vexp\ in the studied GPs.  
The failure to correctly fit troughs in GP\,1133 and GP\,1219, 
where \vexp\ was a free parameter, indicates that \vexp\ alone is not 
sufficient to solve the problem.

\subsection{\lya\ source modelled by a double Gaussian}
\label{sec_2G}

\begin{table}[t]
\begin{center}
\caption{Parameters of double-Gaussian fits of the SDSS \hb\ line.}
\label{tab_2G}
\small
\begin{tabularx}{0.35\textwidth}{ccc}
\toprule
\toprule
ID  & $FWHM_\mathrm{broad}$ [\kms] & $f_\mathrm{broad}$/$f_\mathrm{narrow}$ \\
    & (1) & (2) \\
\midrule
GP\,0303 & 400 $\pm$ 100 & 0.20  $\pm$     0.05\\
GP\,0816 & 350 $\pm$ 100 & 0.  \\  
GP\,0911 & 450 $\pm$ 100 & 0.5  $\pm$      0.1\\
GP\,0926 & 500 $\pm$ 100 & 0.50  $\pm$     0.03\\
GP\,1054 & 500 $\pm$ 100 & 0.33  $\pm$     0.07\\
GP\,1133 & 450 $\pm$ 100 & 0.1   $\pm$     0.1\\
GP\,1137 & 500 $\pm$ 100 & 0.30  $\pm$     0.05\\
GP\,1219 & 500 $\pm$ 100 & 0.33  $\pm$     0.05\\
GP\,1244 & 400 $\pm$ 100 & 0.25  $\pm$     0.05\\
GP\,1249 & 350 $\pm$ 100 & 0.2   $\pm$     0.1\\ 
GP\,1424 & 500 $\pm$ 100 & 0.20  $\pm$     0.03\\
GP\,1458 & 400 $\pm$ 100 & 0.10   $\pm$    0.03\\
\bottomrule
\end{tabularx}
\tablefoot{
(1) $FWHM$ of the broad \hb\ component obtained from double-Gaussian fits,
corrected for instrumental dispersion. 
(2) Ratio of fluxes in broad and narrow SDSS Balmer line components, resulting
from double-Gaussian fits; 
the values represent the mean derived from \ha\ and \hb. 
}
\end{center}
\end{table}

\begin{figure}[t!]
\begin{tabular}{c}
\includegraphics[width=0.45\textwidth]{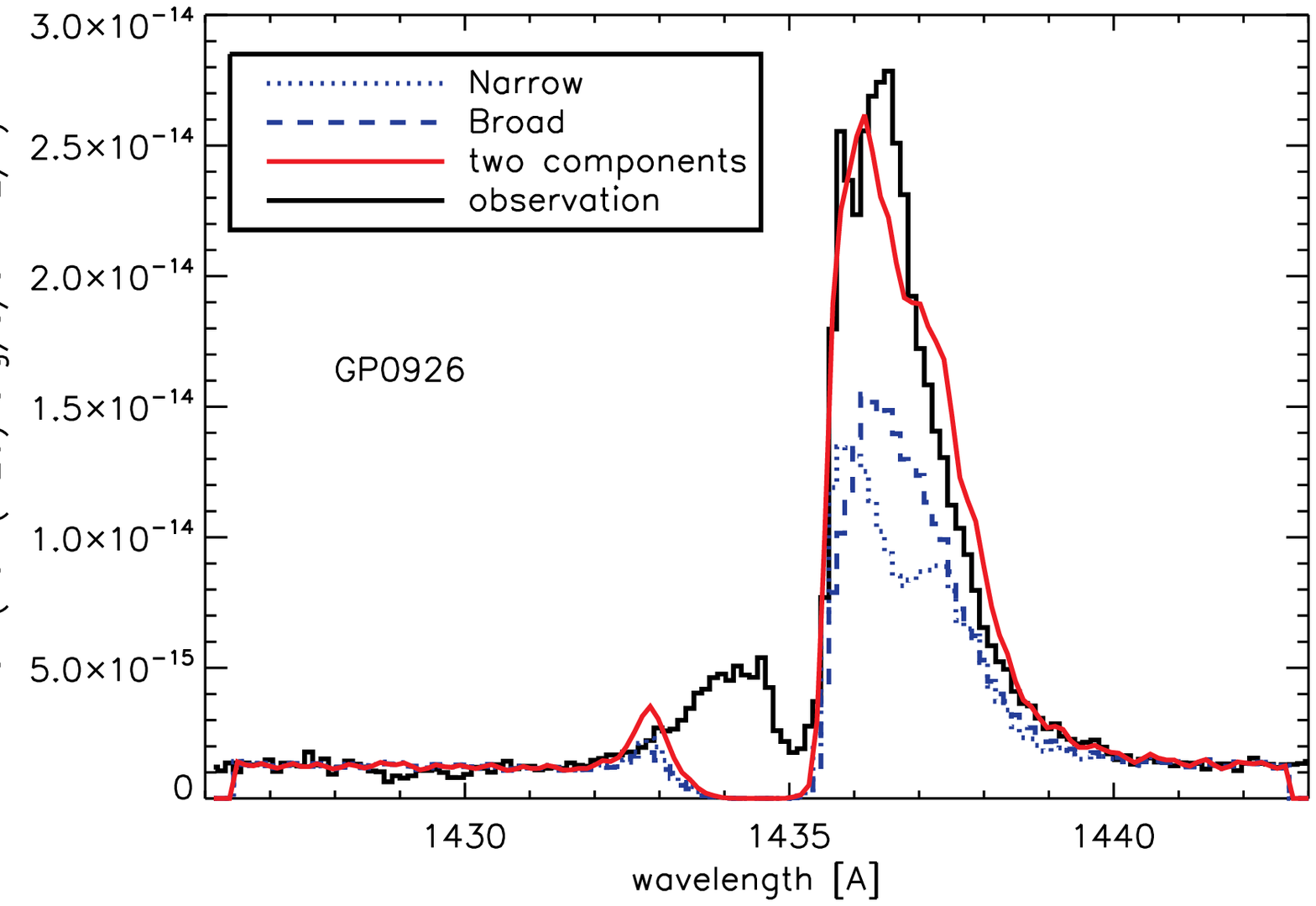}
\\
\includegraphics[width=0.45\textwidth]{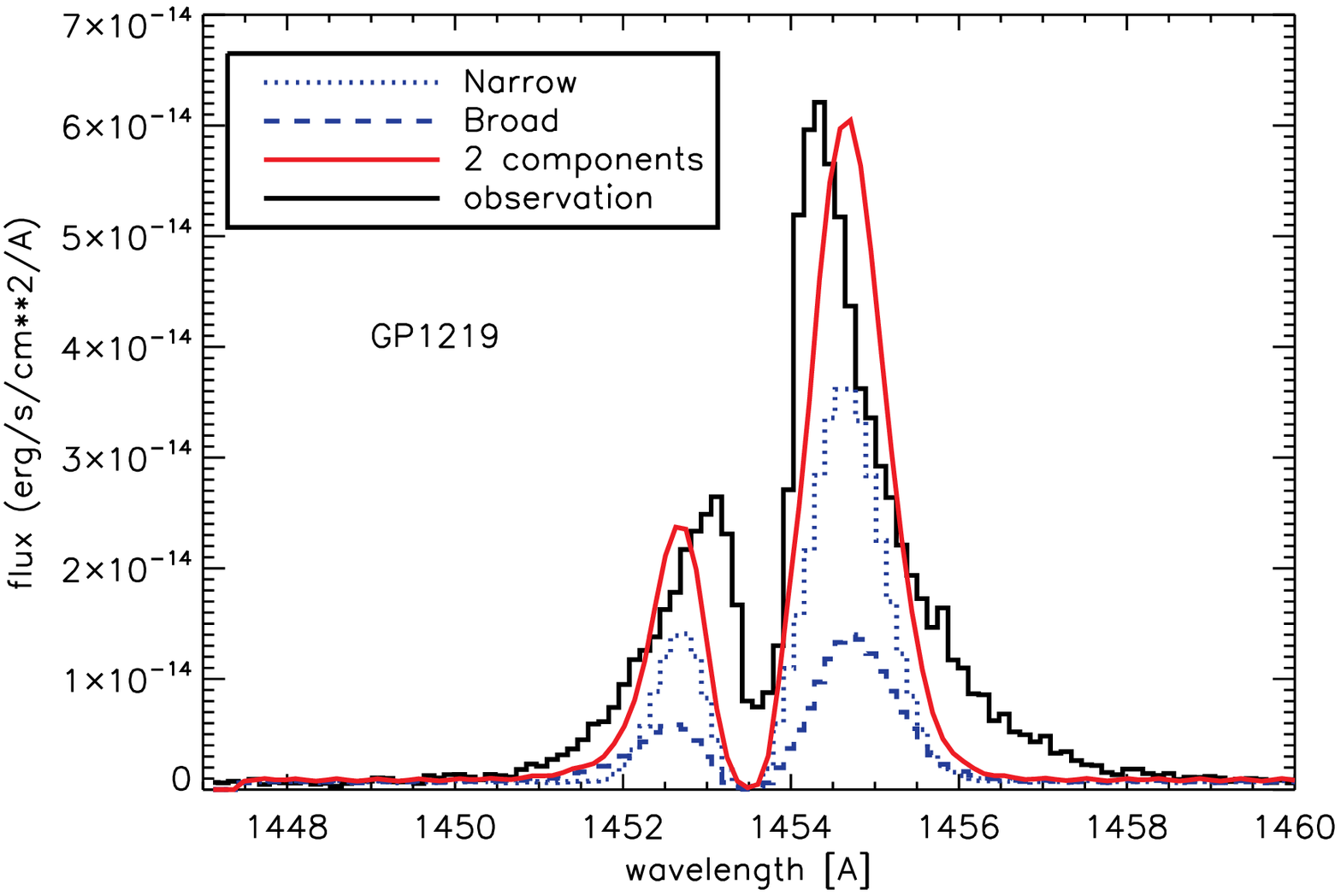}
\\
\end{tabular}
\caption{Constrained fits using double-Gaussian intrinsic \lya\ profiles, 
with a realistic proportion of broad and narrow components, derived from SDSS 
\hb\ spectra. 
Targets with the largest broad component contribution are shown. 
}
\label{fig_2G}
\end{figure}

We searched for the reasons for the bad correspondence between the observed and 
modelled \lya\ profiles resulting from model fitting with constraints.
We first considered a better description of the intrinsic \lya\ profiles
with several kinematic systems. This effort was motivated by 
\citet{Amorin12}, who showed  
complex kinematic structures in GP \ha\ profiles observed  
with a high-resolution ($R\!>\!10\,000$) echelle spectrograph.  
None of their targets have been observed in \lya, therefore no direct 
proof of the effect of multiple kinematic components on the resultant \lya\   
is available. At the SDSS resolution, 
no strong secondary line components are detected in our sample, 
but broad \ha\ and \hb\ wings are consistently seen \citep{Henry15}.

We therefore fitted the SDSS \ha\ and \hb\ lines with double-Gaussian profiles,
narrow and broad. 
At the SDSS resolution, the double-Gaussian fits produce degenerate
results, we  therefore 
used both \ha\ and \hb\ to minimize the degeneracy. The kinematic
parameters were tied between the lines, while they were independent 
between the broad and narrow components. The Gaussian amplitudes 
were free parameters in each line, and we computed the mean between 
\ha\ and \hb\ for each component.  
We found that the broad and narrow components have redshift differences  
$\pm30$\,\kms, that is, they are the same within the uncertainties.  
Unlike in dusty galaxies, the red wing is not extinguished, 
and the broad component is symmetric about the systemic redshift. 
We obtain broad-to-narrow flux ratios in the range $0\!-\!0.5$.
We included the narrow component $FWHM$ in Table~\ref{tab_params}, and  
we list the broad component $FWHM$ and the flux ratios of the broad and narrow 
components in Table~\ref{tab_2G}.   
The flux ratios represent the mean between \ha\ and \hb.
We assume the uncertainty of the broad component
$FWHM$ to be $\sim\!100$\,\kms, 
based on multiple trials of the double-Gaussian fitting. 
All $FWHM$s have been corrected for instrumental dispersion.

We assumed that both narrow and broad \hb\ components were
produced by recombination, and we
used the double-Gaussian profiles as input 
characterizing the intrinsic \lya\ profiles for \lya\ model fitting. 
We present the fitting results for two targets that are among those  
with the largest broad component flux fraction  (GP\,0926, GP\,1219) 
in Fig.\,\ref{fig_2G}.
The resulting fits have not significantly improved over those with single Gaussians. 
Even though the blue peaks have become slightly more pronounced,  
the \lya\ profiles are not well reproduced and the 
problems encountered in Sect.\,\ref{sec_constrained} persist.

\subsection{Unconstrained fitting}
\label{sec_unconstrained}

\subsubsection{Good fits, discrepant parameters}
\label{sect_unconstrained}

\begin{figure*}[th]
\includegraphics[width=1.0\textwidth]{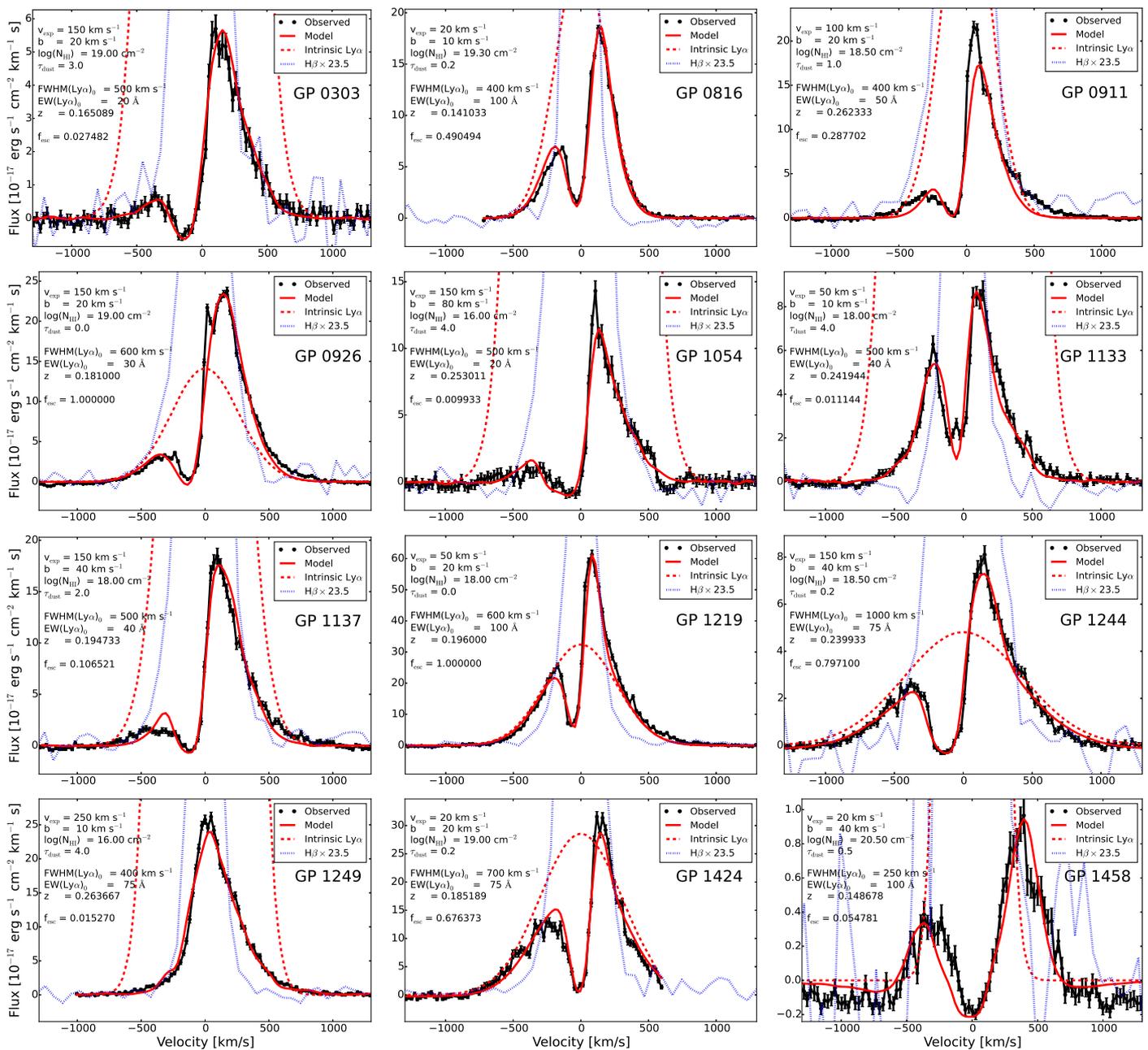}
\caption{Best-fitting \lya\ models, obtained by unconstrained fitting. 
COS data and their errorbars 
are shown with the black line, best-fitting model with the solid red line, intrinsic \lya\ 
profile with the dashed red line, and the observed SDSS \hb\ scaled by Case B factor 23.5 
with the dotted blue line.
Derived best-fit parameters of the shell are shown in each panel. 
 }
\label{fig_fits}
\end{figure*}

\begin{figure*}[th]
\begin{tabular}{lll}
\includegraphics[width=0.31\textwidth]{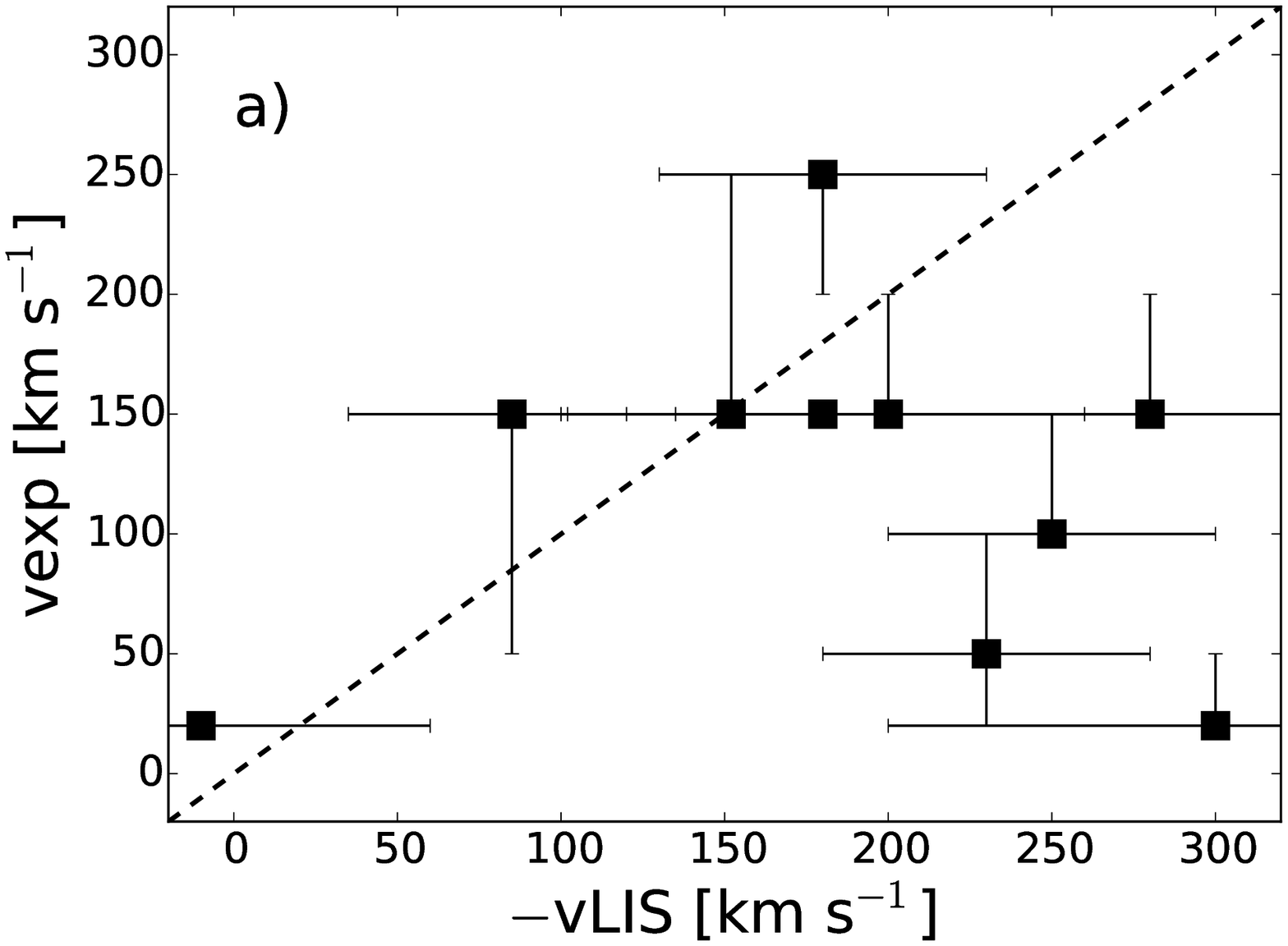}
&
\includegraphics[width=0.31\textwidth]{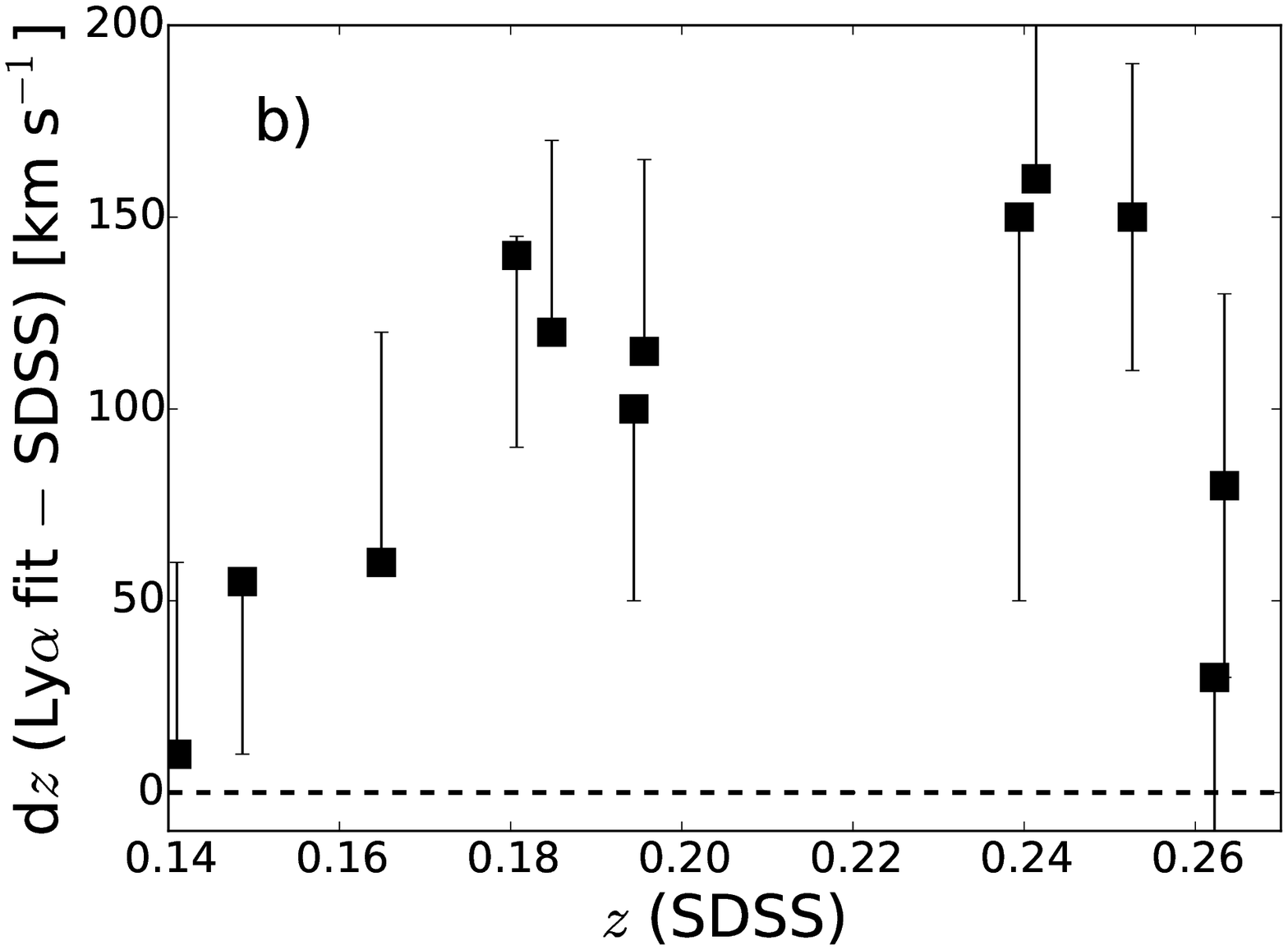}
&
\includegraphics[width=0.31\textwidth]{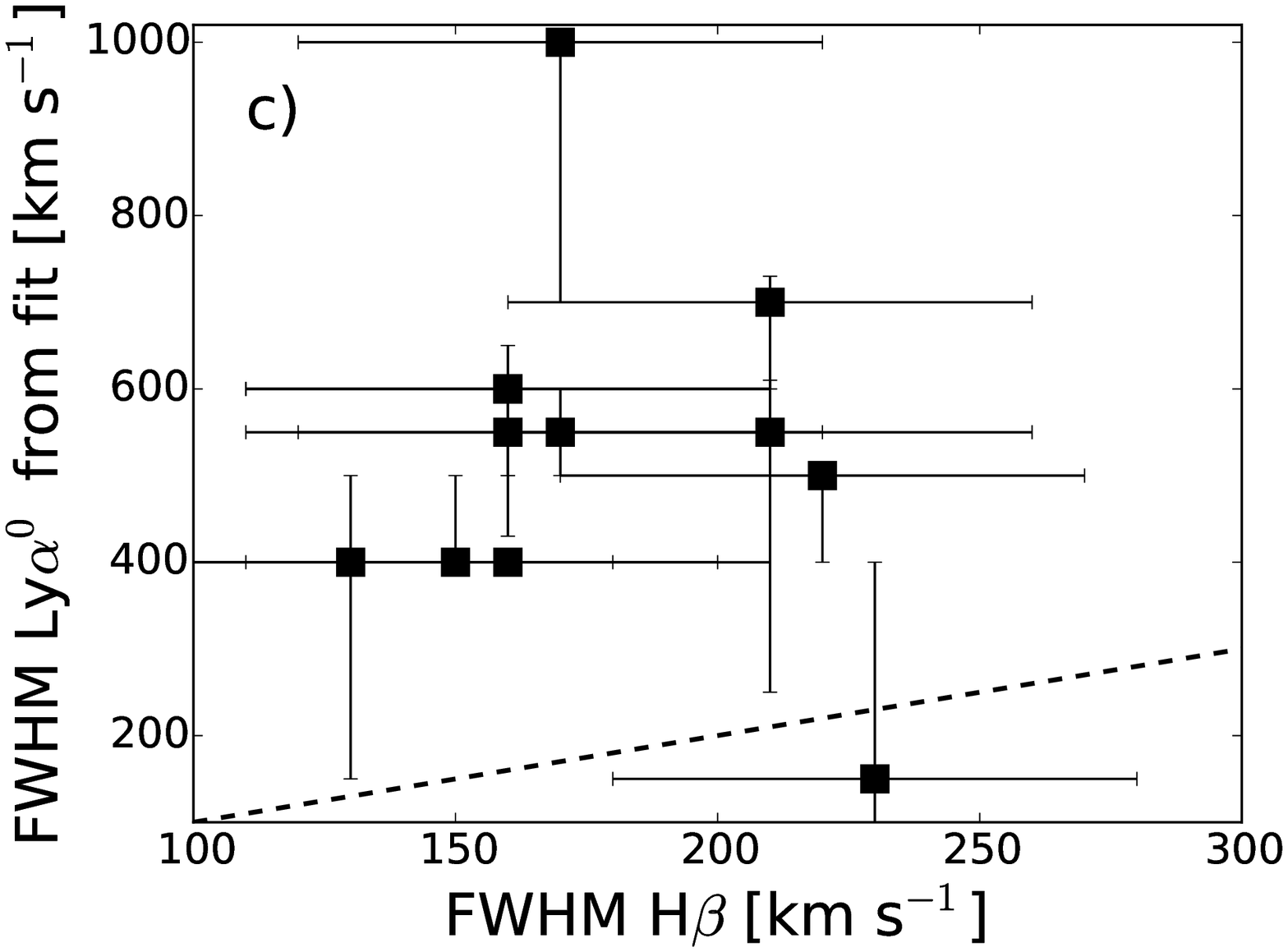}
\\
\end{tabular}
\caption{Comparison between parameters derived from the unconstrained \lya\ 
fitting, and their observed counterparts. a) outflow velocity; b) 
redshift; c)   
\fwhmlyai. The dashed lines show the expected identity between the 
parameters. The points and errorbars correspond to those listed in 
Table~\ref{tab_params}.
}
\label{fig_parameters}
\end{figure*}

To understand which parameters play a role in producing 
the unsatisfactory results of the 
constrained model fitting,  
we finally relaxed all the constraints. 
We ran the fitting process across
the entire grid of $>\!6000$ {\tt MCLya} shell models (with four varying 
shell parameters \vexp, $b$, \nh, and $\tau_\mathrm{d}$), and we ignored the 
constraints from the ancillary data. In addition, the parameters of the \lya\ 
source, that is, \ewlyai, \fwhmlyai, and $z$, were also let free. 
A good match between the best-fitting modelled and observed spectra has 
been found by the automatic fitting procedure 
for all of the twelve targets, unlike in \citet{Yang16}, where fitting 
of GP\,1133, GP\,1219 and GP\,1424 failed. 

We present the lowest $\chi^2$ model for each galaxy in Fig.\,\ref{fig_fits}.
To assess how the model parameters differ from the 
observations, we considered twenty lowest-$\chi^2$ models for each target.
A visual check revealed that for the given grid resolution, 
the twenty models encompassed 
spectra that were still in reasonable agreement with the observations: they matched the positions and amplitudes of the \lya\ peaks and troughs, 
and matched the line wings. A larger number of fits was inappropriate
due to the deteriorating fit quality, and 
we did not opt for a narrower set in order to allow for possibly 
large intervals in all the 
fitting parameters, while searching for overlaps with the
observed ones.   
The visual check was complementary to the $\chi^2$ parameter computation. 
This simple method 
 allowed an assessment of the parameter range without a complex fit quality
analysis, unnecessary for the problem in hand. 
To characterize the fitting parameter distribution in each target, 
we computed their median and the $10^\mathrm{th}$ 
and $90^\mathrm{th}$ quantiles in the set
of twenty best fits, and used them 
in Table~\ref{tab_params}, columns 6\,--\,12 (the uncertainties here 
correspond to the quantiles). 
The quantiles are convenient for the discrete 
character of the grid and our choice discards only the extremes in each given 
set.

Figure\,\ref{fig_fits} shows that the homogeneous shell models are 
able to reproduce the GP \lya\ spectra,  as in 
\citet{Yang16}. 
However, a comparison between the observed and 
modelled ISM parameters reveals 
significant disagreement (Table~\ref{tab_params}). 
We have identified three major 
discrepancies, which we describe below and illustrate in 
Fig.\,\ref{fig_parameters}:

\begin{enumerate}
\item
The only way to produce double-peaked \lya\ profiles in the shell models is 
by assuming 
a static or low-velocity ($\lesssim\!\!150$\,\kms) \hi\ medium 
\citep[see][]{Neufeld90,Verhamme06,Dijkstra06}. 
Therefore, except for the single-peak target GP\,1249, 
all the best fits of our sample have  
\vexp\,$\leq$\,150\,\kms. Spectra with the  
strongest blue peaks (GP\,0816, GP\,1133, GP\,1219, GP\,1424, and 
GP\,1458) required even lower velocities, \vexp\,$\leq$\,50\,\kms. 
In contrast, the observed LIS velocities \vlis\ are significantly larger in 
at least one third of the sample (Fig.\,\ref{fig_parameters}a).
No convincing LIS line detection was possible in GP\,1133 and GP\,1219, but 
their high-ionization gas velocities 
from \ion{Si}{iii} and \ion{Si}{iv} are between  
$-300$ and $-400$\,\kms, and we also find \lyb\ components at similar 
velocities, inconsistent with the fitted \vexp\,$\sim50$\,\kms. 
We list the measured LIS velocities with a negative sign in Table\,\ref{tab_params} 
(the lines are blueshifted), 
while we provide a positive \vexp, which is defined as the expansion velocity of the shell. 
\item
Systemic redshifts derived from the \lya\ models are larger than those 
from the SDSS emission lines in every target of the sample.   
With the exception of three targets, the redshift discrepancy $\Delta z$ 
is larger than the 
conservative wavelength calibration uncertainty of 40\,\kms, and reaches values as high
as $\sim\!\!160$\,\kms\ (Fig.\,\ref{fig_parameters}b). 
The targets with redshifted \lya\ troughs 
(e.g. GP\,1133, GP\,1424) described in Sect.\,\ref{sect_profiles} do not stand out and 
are part of the trend.
The best-fitting model redshifts 
are driven by the requirement that the central trough should lie at the 
position of the largest optical depth, close to $-$\vexp. 
\item
Large intrinsic \lya\ line widths, \fwhmlyai, were chosen by the best-fitting 
models to match the shape of the blue \lya\ peaks, 
and the broad blue and red wings. These 
compensate for the lack
of profile broadening in low-\nh\ models, determined by the observed
small separation of \lya\ peaks.
The typical best-fit \fwhmlyai\,$\sim$\,500\,\kms\ 
(Fig.\,\ref{fig_parameters}c) are   
much larger than the measured SDSS \fwhmhb\,$\sim$\,200\,\kms, and are  
similar to or larger than the broad component of \hb. 
However, we cannot conclude from here that the narrow component was attenuated
and only the broad one was transmitted.  
The modelled intrinsic \lya\ lines have larger fluxes than the  
broad components of Balmer lines scaled by the Case B factors. Therefore, 
we need to interpret this discrepancy as another model failure, not a physical effect.   
\end{enumerate}

The remaining free fitting parameters, $b$, \nh, $\tau_\mathrm{d}$ and \ewlyai,
 either have no directly 
observable counterparts or their comparison to observed values is less
straightforward.  
No measurable counterpart was available for $b$
and \nh. 
The Doppler parameter $b$ is the least robust fitting parameter 
\citep[see also][]{Gronke15}, therefore its large spread
in some of the targets is unsurprising.   
On the contrary, \nh\ is among the most robust parameters 
due to its strong impact on the \lya\ peak shift 
\citep{Verhamme06,Gronke15}. The \nh\ that we here derive from the shell models
is generally low ($\lesssim10^{19}\mathrm{\,cm}^{-2}$) 
compared to typical star-forming galaxies.  
The modelled FUV attenuation due to dust, $\tau_\mathrm{d},$ 
has a large scatter for each target. 
 This is most probably caused
by the low \nh. 
The role of dust in low-\nh\ environments is reduced due to a relatively
small number of scattering events and therefore 
models with a small and large dust content thus give similar 
results. Nevertheless, given the large uncertainties in both the  
modelled and observed $\tau_\mathrm{d}$ 
(due to the low \nh\ and the uncertainty in the FUV attenuation 
law), we cannot properly judge the consistency between the observations and the 
models: we can only state that the results are consistent 
within the large error bars. 
Finally, the best-fitting model \ewlyai\ is usually 
lower than that derived from \hb\ 
observations in the studied sample, or is at the lower edge of the interval 
of possible \ewlyai\ 
values (Table\,\ref{tab_params}). This may indicate that a part 
of \lya\ has been transferred outside the COS 
aperture, which is consistent with the result obtained from a comparison 
of the COS and GALEX observations for two GPs
\citep{Henry15}, and from the \lya\ HST imaging for GP\,0926 \citep{Hayes14}. 
It can also indicate a preference for extinction curves with a low total to 
selective extinction ratio, 
typical of low-metallicity galaxies and also found in GP-like galaxies 
of \citet{Izotov16,Izotov16b}.

\subsubsection{Fit parameter discrepancies are tied to spectral shape}
\label{sec_ewbr}

We have tested whether any correlations exist between the fitting parameters
or their discrepancies. 
The first conclusion is that both \vexp\ and $z$ need to be free parameters
in order to reproduce the GP \lya\ peaks and troughs, but nothing can be 
concluded about the correlation between the discrepancies 
$\Delta z$ and $\Delta v$.  
Secondly, a hint of correlation was found between the 
redshift discrepancy $\Delta z$ and the ratio of the 
intrinsic \lya\ and \hb\ widths, \fwhmlyai/\fwhmhb.
Larger samples are needed to confirm or refute the correlations.

\begin{figure}[th]
\begin{tabular}{l}
\includegraphics[width=0.45\textwidth]{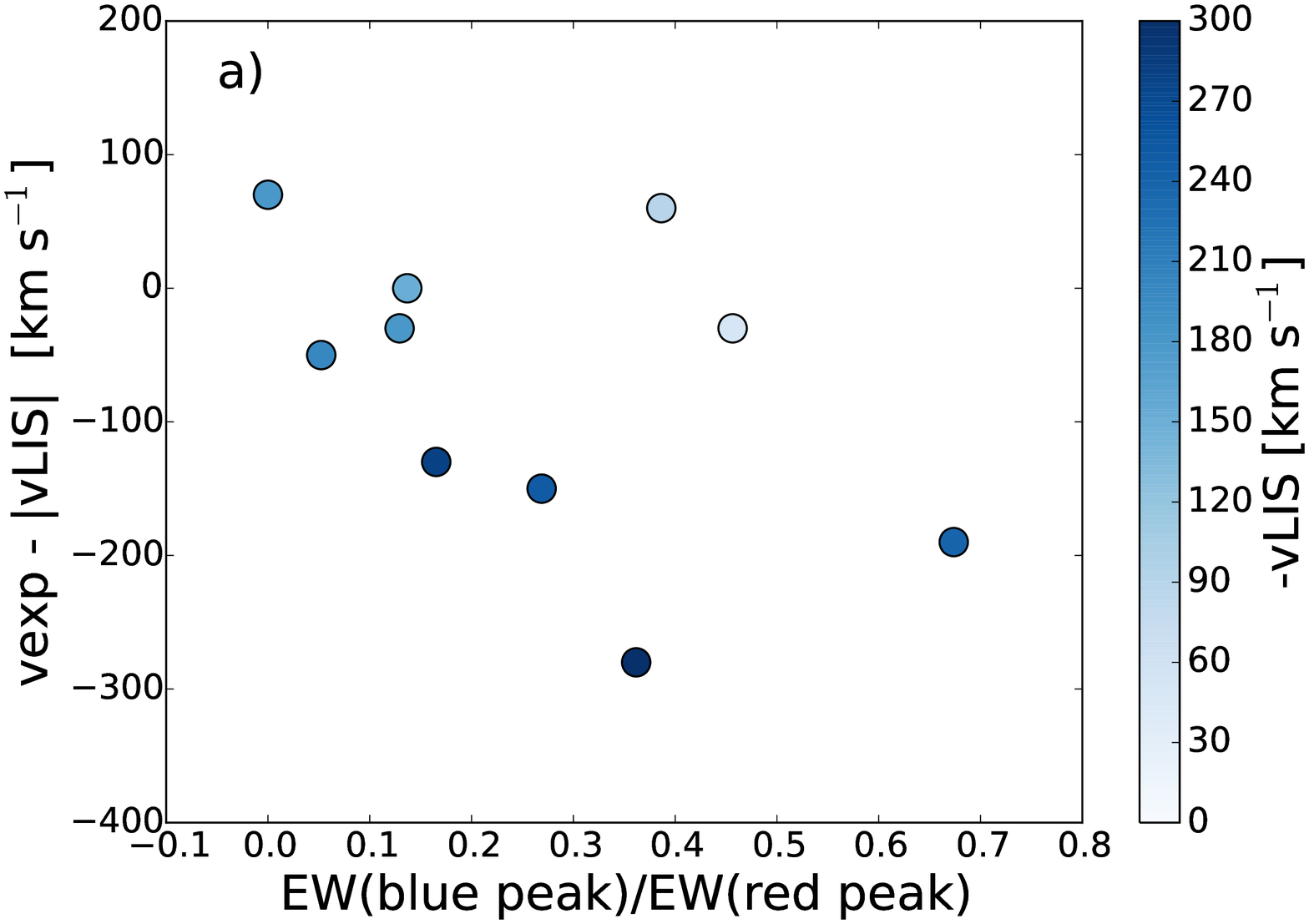}
\\
\includegraphics[width=0.45\textwidth]{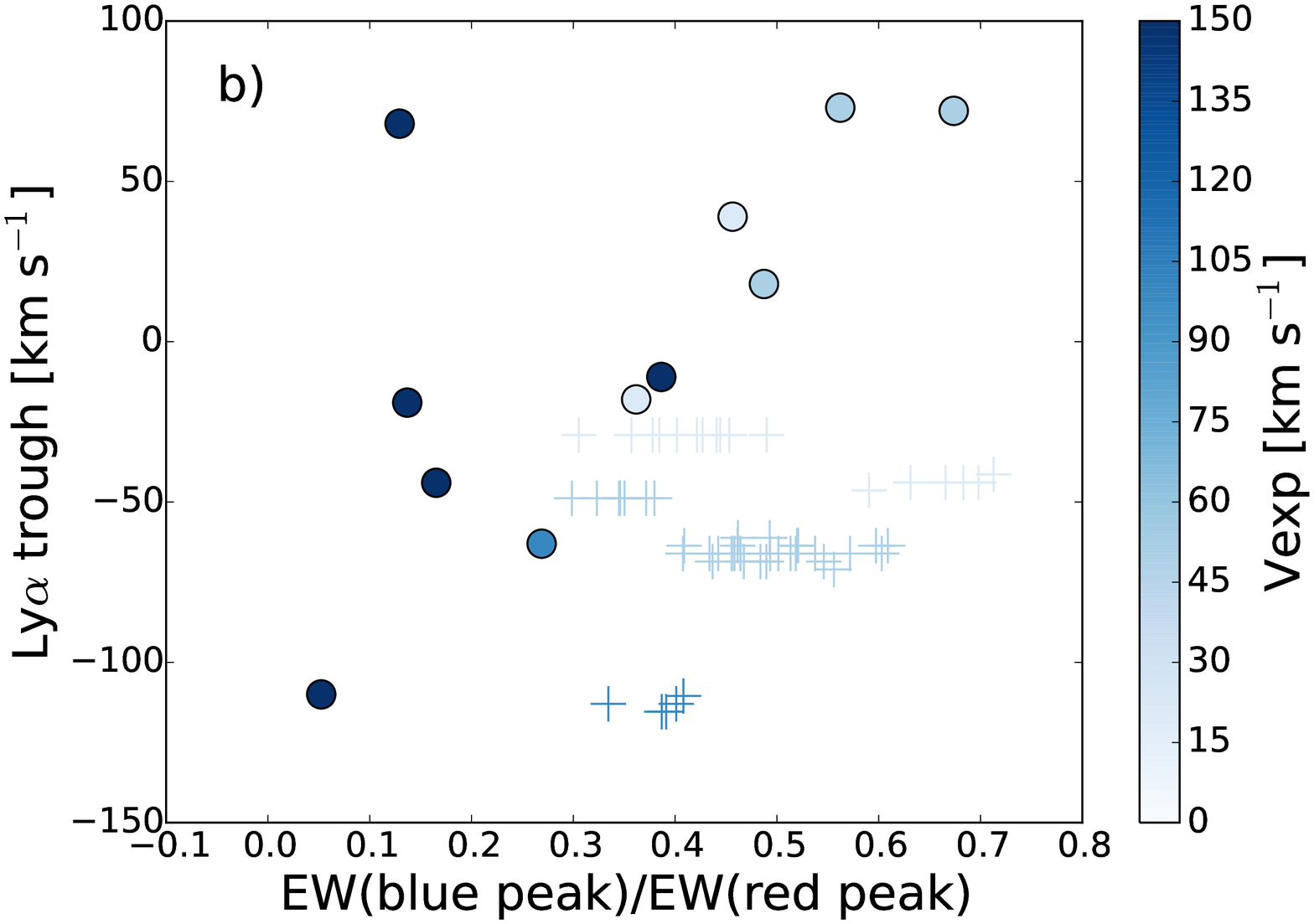}
\\
\includegraphics[width=0.45\textwidth]{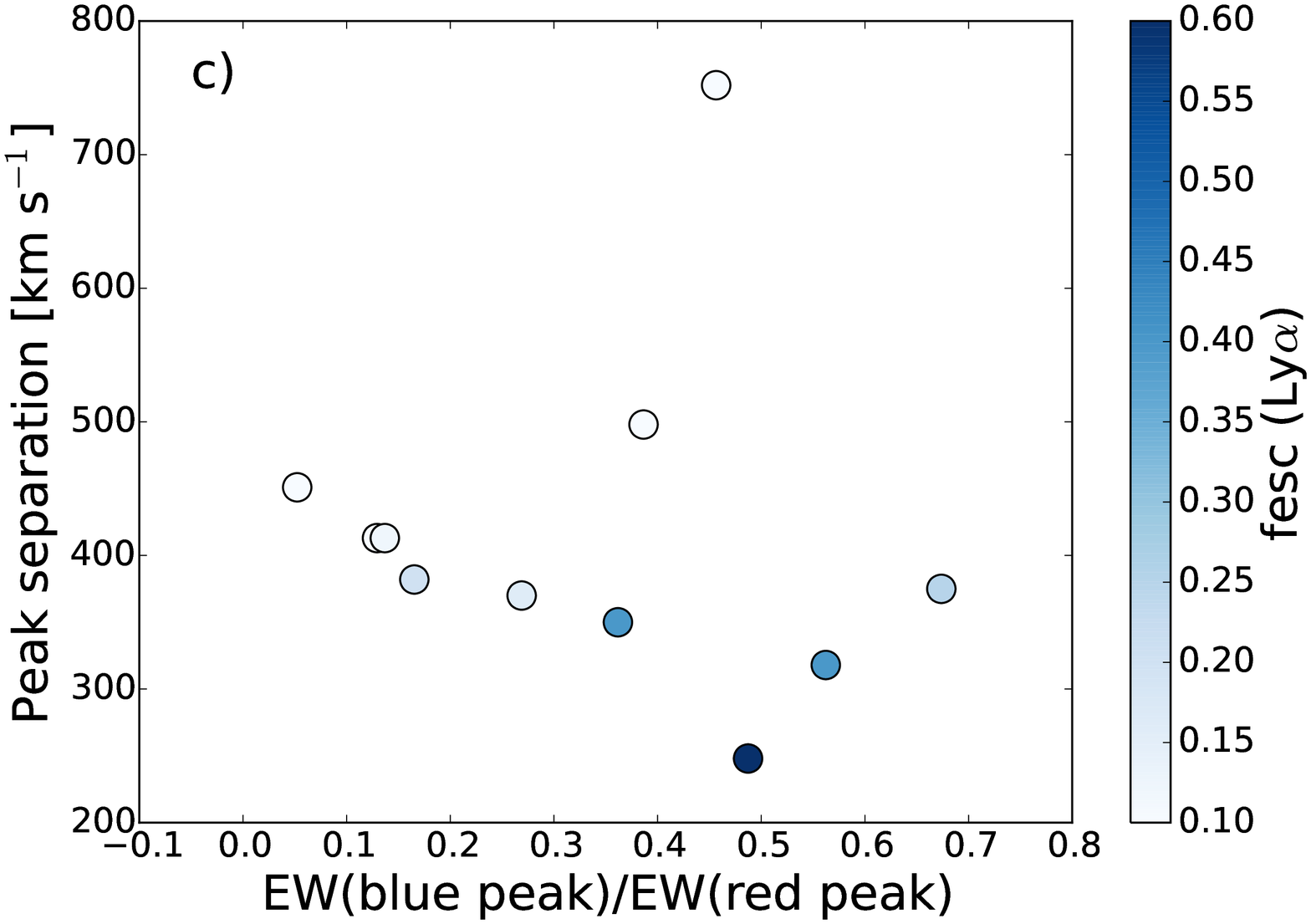}
\\
\end{tabular}
\caption{Relations between the \lya\ blue and red peak $EW$ ratio, and several
observed and best-fit parameters. Circles represent the observed GPs, crosses
represent shell models. Details can be found in Sect.\,\ref{sec_ewbr}.  
}
\label{fig_ewbr}
\end{figure}

A natural question is how the difficulties in 
\lya\ fitting are linked to the double-peak character of the line profile, 
common in the GPs. We devote Fig.\,\ref{fig_ewbr} to studying 
several best-fit and observational parameters as a function of the 
\lya\ blue peak flux fraction, expressed as \ewlyab/\ewlyar.  
The blue and red peak fluxes were measured as 
separated by the central trough (and not by the systemic 
redshift) to better characterize flux in each peak,  
described in Sect.\,\ref{sect_profiles}. 
We find that 
the difference between fitted and measured gas velocities, \vexp$-$|\vlis|,   
generally increases with the increasing blue peak strength 
(Fig.\,\ref{fig_ewbr}a). 
With the exception of two points that have 
high \ewlyab/\ewlyar$\sim0.4$ and \vexp$-$|\vlis|$\sim0$, all the other
studied galaxies 
show an anti-correlation between the velocity difference and the blue peak
flux fraction (Spearman coefficient $-0.91$ and P-value 0.002).
The colour-scale coding of the plot 
helps clarify that the two outliers have low LIS velocities 
$\left(<\!100{\textrm{\,\kms}}\right)$. Thanks to them, there is a
fortuitous agreement
between the modelled and observed values, due to the fact that 
low velocities are the only way to produce strong blue peaks in the shell
models. 
The colour scale of Fig.\,\ref{fig_ewbr}a also helps to visualize that
the double peaks appear in GPs across a large range of LIS velocities. 
If the GP \lya\ blue peaks were due to low gas velocities, 
we would expect a correlation between the LIS velocities and the
observed \ewlyab/\ewlyar\ -- which we cannot confirm in our data.
The blue peaks 
may thus not be linked to static environments, or alternatively may be
formed in static environments not probed by the LIS lines.

Figure~\ref{fig_ewbr}b is devoted to the \lya\ trough 
position, which we described earlier in this paper to be 
inconsistent with the measured LIS velocities. 
We observe that as the blue peak grows stronger, 
the trough shifts from large negative offsets towards 
the systemic redshift and then to positive offsets.  
The peculiar shift to positive velocities, unexpected from the modelling, 
 is just a continuation of the overall trend. 
We also see a trend between the trough position and \vexp\ (colour scale in the
plot): the trough position is most negative for largest \vexp\ ($>\!100$\kms), 
and positive mostly for low \vexp. 
A correlation between the trough position and both \vexp\ and relative 
blue peak $EW$ 
is indeed expected in the models where the blue peak originates from
radiative transfer in low-velocity or static environments:  
the optical depth is maximum at approximately $-$\vexp, therefore the 
trough offset from the systemic velocity becomes larger with increasing \vexp. 
The blue peak becomes weaker with increasing \vexp. 
However, the models (cross symbols) are not co-spatial with the observed data
in Fig.\,\ref{fig_ewbr}b.
The model troughs are shifted further to negative velocities 
and can never reach positive velocities. 
This illustrates why an adjustment of the systemic redshift is needed to fit
the observed profiles.

Finally, Fig.\,\ref{fig_ewbr}c
shows the \lya\ double-peak separation as a function of the blue-to-red $EW$ 
ratio and the observed \fesclya. 
Even though purely observational, this plot elucidates the reasons
for the incompatibility between the models and GP data.     
The diagram shows that the stronger the blue peak, the smaller 
the peak separation in most cases, and  
the larger the observed \fesclya\ (colour scale in the same plot).
The two outliers with peak separations 
of $\sim\!500$\,\kms\ and $\sim\!800$\,\kms\ are the same ones as in 
Fig.\,\ref{fig_ewbr}a, with 
nearly static kinematics that favour the blue peak formation.  
The remaining nine GPs 
do not seem to have blue peaks related to the outflow kinematics 
but rather to the low \hi\ column density for the following reasons: 
1) \lya\ double-peak separation is a good tracer of the \hi\ column density
\citep{Verhamme15,Verhamme17}. As we find the smallest \lya\ separation 
in GPs with the largest blue peak flux contribution 
(Fig.\,\ref{fig_ewbr}c), the probability of 
the LyC escape increases with the increasing blue peak flux. 
2) LyC escape from the GP-like galaxies has been shown to correlate with 
\fesclya\ \citep{Verhamme17}. 
In our sample, GPs with a large blue peak tend to have a larger 
\lya\ escape fraction 
(colour scale in Fig.\,\ref{fig_ewbr}c). 
The anti-correlation between the \ewlyab/\ewlyar\ 
ratio and \fesclya\ has the Spearman coefficient of 0.85 and p-value of 0.003 
if the two outliers with static LIS gas are left out. 
In addition, we see an 
anti-correlation between the peak separation and \fesclya, with 
the Spearman coefficient of 0.94 and p-value $10^{-4}.$ 
Therefore, if the blue peak is not produced
in static gas, it is related to the \lya\ and potentially the LyC escape.  
In complementarity to 
previous GP studies \citep{Henry15,Yang16,Yang17} we therefore 
show here that not only the blue peak position but also its relative flux 
seems to carry information about the transparency of the neutral ISM.

\section{Discussion}
\label{sec_discussion}

\begin{figure}
\includegraphics[width=0.48\textwidth]{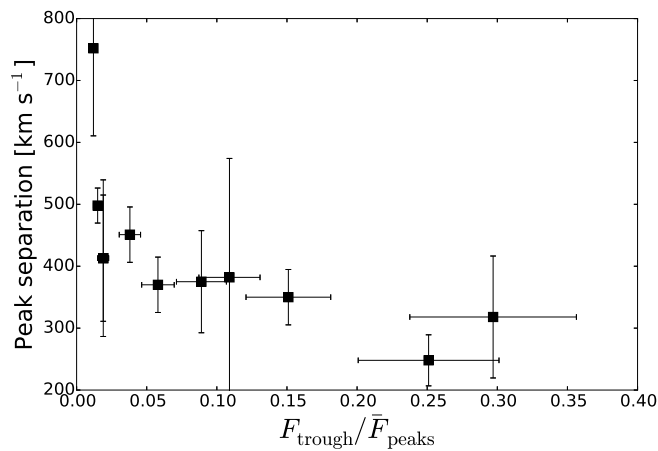}
\caption{\label{fig_trough_sep}
Observed separation between the red and blue \lya\ peaks 
as a function of the  
residual flux in the central \lya\ trough $F_\mathrm{trough}$
divided by the mean flux in the red and blue \lya\ peaks  
$\bar{F}_\mathrm{peaks}$. 
An analogous plot was presented for model spectra in \cite{Gronke16}.
No correlation was seen for clumpy models, while a trend was present for 
homogeneous shell models (see Sect.\,\ref{sec_clumpy}).
}
\end{figure}

We devote the following sections to the exploration of possible reasons 
for the encountered discrepancies, 
discussion of the possible \lya\ origin, 
and the compatibility of the observations with other existing models.
We have searched for correlations between the fit parameter discrepancies 
and galaxy properties such as mass, size, metallicity, 
star-formation rate, amount of dust, UV absorption line $EW$s, 
emission, and absorption line $FWHM$. We found no clear trends. 

\subsection{Blue peaks in the literature}

The unconstrained \lya\ fits that we presented in 
Sect.\,\ref{sect_unconstrained} were previously studied by \citet{Yang16}, 
using the same GP sample and   
a similar set of homogeneous shell models  
\citep[][]{Dijkstra06,Gronke15}. 
In this respect, the present paper reproduces their fitting results, while it 
further extends the analysis by applying observational constraints and by  
measuring the differences between the modelled and observed ISM parameters.
In this paragraph, we
compare the two sets of unconstrained fitting results produced by the
two papers.  Unlike the present paper, 
\citet{Yang16} did not convolve the synthetic
spectra to the observed spectral resolution, which could be the origin of 
several discrepancies. 
While our automatic procedure fitted all of the twelve \lya\ profiles, 
\citet{Yang16} were able to fit nine of them 
(their Figure 7). They reported that they manually adjusted 
the model parameters for the remaining three 
 -- GP\,1133, GP\,1219, and GP\,1424 -- to match the observed peaks and 
troughs. They published the model parameters for all twelve galaxies 
in their Table~4, which we now compare with our Table~\ref{tab_params}.
The derived shell expansion velocities were similar in both studies 
(with differences between them of $\lesssim$\,50\,\kms), 
and are equally inconsistent with the measured LIS outflows. 
Both studies needed similarly broad intrinsic \lya\ profiles 
to achieve good fits
(agreement between the two codes within $\sim$\,100\,\kms).
The third parameter causing problems in our fitting, the 
systemic redshift, was not discussed by \cite{Yang16}. They 
only presented the SDSS redshift in the paper and did not discuss 
any adjustments, 
therefore a comparison with our results cannot be done. 
No data were given for \ewlyai\ in their paper either. 
Comparison between the derived \hi\ column densities 
shows that in approximately one half of the sample 
the values agree between the two studies. In the other half, 
we typically find \nh\ lower by $\sim\!\!1/2$\,dex. 
We speculate that the reason lies in the absence of 
correction for spectral resolution in \citet{Yang16}; 
instead of broadening by the instrumental effect, 
their synthetic profiles needed to be additionally broadened by a higher \nh. 
For the dust optical depth $\tau_\mathrm{d},$ both papers 
show a large scatter in the best-fitting values for each target. 
The reason is a weak effect of the dust on \lya\ 
in low-\nh\ models (\nh\,$\lesssim\!\!10^{19}$\,cm$^{-2}$), which 
are characteristic of the GPs and which allow an efficient
\lya\ escape. Both dust-poor and dust-rich \hi\ media can produce 
similar \lya\ spectra if the \nh\ is low, and are thus equally likely 
to be among the best-fitting models. 
Our median $\tau_\mathrm{d}$ values were lower by 
$\Delta\tau_\mathrm{d}\!\sim\!0.3\!-\!1$ than those of \citet{Yang16}  
in one half of the sample, while 
they were either equivalent or higher by a similar amount in the other half. 
Given the large spread of $\tau_\mathrm{d}$ for each target, the values can be 
considered consistent between the two studies (with the exception of GP\,0911), 
and also in agreement with the observations within the error bars. 
As for the Doppler parameter $b$ 
\citep[which corresponds to the temperature in][]{Yang16}, 
it is the least robust among the fitting parameters
\citep{Gronke15}, therefore a large scatter in the fitted values
and large differences between the two codes would not 
be surprising. We see from the \citet{Yang16} results that 
their model grid 
admitted temperatures as low as $10^3$\,K, which was an order lower 
than in our code. We therefore automatically obtained fits with higher
temperatures.   
In summary, we consider the results of the two unconstrained-fitting 
studies to be mostly consistent within the error bars.  
We generally considered larger error bars 
in order to account for the discreteness of the model grid
and for the observational uncertainties, and 
also to provide sufficient parameter space for testing 
the match between the shell models, the ISM parameters, 
and the observational data.

\begin{figure*}[t!]
\includegraphics[width=1.0\textwidth]{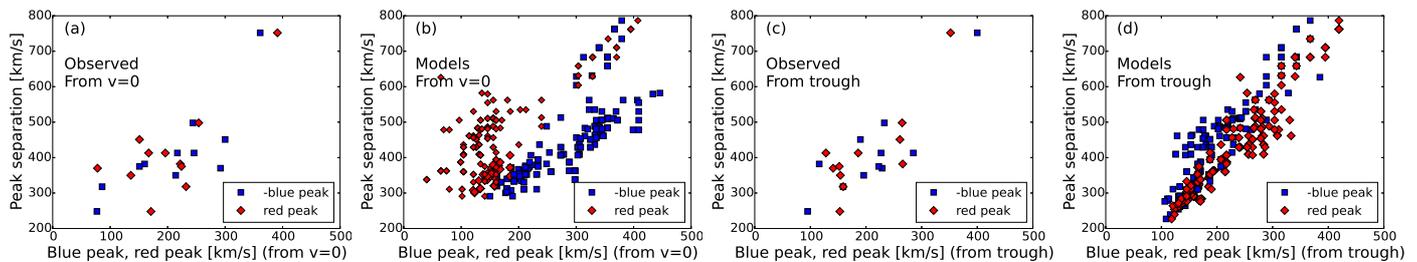}
\caption{Symmetry of the red and blue \lya\ peak positions in the observed and 
modelled spectra. 
a) The observed \lya\ double-peak separation as a function 
of the blue and red \lya\ peak positions: 
the distributions are similar for both peaks, 
the correlation is tighter for the blue peak, 
b) The same as (a) but for shell models; here  
the correlation is tighter for the blue peak, and   
the red and blue peaks are not symmetric.  
c) The same as (a) but measured with respect
to the central \lya\ trough; the correlations here are stronger than in (a). 
d) The same as (b) but measured with respect to the central \lya\ trough; 
the asymmetry between  the red and blue peaks disappears. 
}
\label{fig_symm2}
\end{figure*}

\bigskip
 
We also note here that problems with the double-peak \lya\ profile fitting 
have been reported 
in the literature before. \citet{Chonis13} had problems fitting double-peak profiles of $z\sim2$
LAEs. However, we believe that those problems were mainly caused by the models that they used. 
Their radiative transfer code only computed the radiative transfer of monochromatic \lya\ radiation, 
unlike ours that assumes a Gaussian line input plus a continuum. We are able to reproduce their LAE \lya\ profiles with our model grid 
in the same way as the LBG spectra in \citet{Verhamme08}, 
\citet{Schaerer08}, and as the GP spectra 
with unconstrained models in this paper. 
The problems that we encounter in our GP fitting are of a different nature: 
due to the availability of the detailed constraints we see the discrepancy between the fitted and observed ISM
parameters. Such a detailed study has only been possible in the low redshift so far. Some of the constraints
were available in the $z\sim2$ LAE study of \citet{Hashimoto15}, 
where they also needed  to invoke broad intrinsic \lya\ to reproduce the 
blue peaks.

\subsection{Stellar \lya}

The double-peaked \lya\ profiles observed in the GPs have their central 
troughs shifted in velocities compared to what was expected from the LIS
absorption lines. In addition, some of the \lya\ troughs are redshifted 
to positive velocities, unexpected in outflowing media.    
We have tested whether, despite the LIS line results, the unusual \lya\ troughs  
could be explained by  
a combination of outflowing and infalling shell models 
The answer was negative. 
Infalling spherical \hi\ shells with the \lya\ source placed in the sphere
centre would produce a redshifted trough, but the line profile symmetry
would be opposite, with the blue
peak dominating over the red \citep{Verhamme06,Dijkstra06}. 
A simultaneous production of 
a dominating red peak (as observed) and a redshifted trough (as observed) 
is challenging. We tested 
several scenarios that superposed infalling and outflowing shells, 
and none of them produced 
a redshifted \lya\ trough for the given flux ratio of the red and blue peaks.  
We conclude here that the redshifted
troughs probably point either to inconsistencies between 
the redshifts probed by \lya\ and \hb, or to 
ISM geometries not covered by our models, such as clumps, which we will 
discuss in Sect.\,\ref{sec_clumpy}. 
Related to this, \citet{Yang16} discussed the possibility of infalling clumps,
but with no direct proof or conclusion.

We have tested the possibility that the 
shift could be due to an underlying stellar \lya\ absorption. 
Synthetic stellar population models using theoretical stellar atmospheres 
\citep{Pena13,Verhamme08,Valls93} showed that for a staburst $<$\,5\,Myr,
the stellar \lya\ reaches absorption \ewlya$_\mathrm{stell}\!\sim\!-4\pm2$\,\AA,
depending on the stellar model and on the star-formation regime. 
Such young starbursts are expected in our sample due to their large \ewha\, 
and were proven by stellar population fitting 
of similar targets in \citet{Izotov16,Izotov16b}. 
Therefore, the stellar \lya\ absorption reduces the observed emission fluxes by 
$<\!\!10\%$ in most of our sample.
Furthermore, the discrepancy of the trough position is the largest
in the strongest \lya\ emitters of our sample (Sect.\,\ref{sec_ewbr}), 
where the \lya\ absorption will represent a particularly low 
fraction of the total flux.
We have nevertheless tested the role of the stellar absorption on the 
final \lya\ line profile, by matching Starburst99 models \citep{Leitherer99} 
of different ages and metallicities to our observational data. 
The synthetic spectra over-predict the \lya\ absorption due to their use of  
observational O-star spectra, contaminated with interstellar features 
\citep{Pena13}. With this in mind, we selected synthetic
spectra that best matched the \ion{N}{v}\,$\lambda1240$ stellar P-Cygni 
feature. We subtracted them from the observed GP \lya\ spectra. 
Even the over-predicted stellar \lya\ absorption did not change the \lya\ 
profile, namely, they did not shift the central trough position.

\subsection{Galaxy compactness and \lya\ halos}
\label{sec_halo}

Numerical studies \citep{Verhamme12, Verhamme15b} have raised the possibility
of \lya\ double-peak formation in the \lya\ ``halos'', that is, in 
extended
diffuse \lya\ emission regions with no corresponding \ha\ or stellar light. 
Radiative transfer in the cited hydrodynamic simulations 
showed that \lya\ spectral profiles varied with 
the aperture size and position. Diffuse \lya\ halos were observationally found
in stacks of high-$z$ galaxies
\citep{Steidel11,Momose14},
and confirmed in individual galaxies both at low and high redshift
\citep{Hayes13,Hayes14,Hayes16,Wisotzki16,Patricio16}.
 
The prevalence of double-peaks distinguishes the GPs from other galaxy 
samples. We can thus hypothesize that we see the effect of the \lya\ halos 
here, similar to the simulations.     
GPs are compact galaxies, with the UV angular size generally
smaller than the HST/COS aperture, unlike in the case of other local galaxies 
where the COS aperture contains the signal from one star-forming knot 
\citep[such as in][]{Wofford13,Rivera15}.
Thanks to the large COS \lya\ escape fractions measured in GPs, 
we expect that \lya\ does not scatter too far from the production sites and  
does not create large halos, and that most of their signal is contained
in the COS spectra. 
However, whether or not the double peaks are due to the non-resolved nature
of GPs and \lya\ halos needs 
to be observationally tested in more detail.  
GP halos have so far been directly observed only for 
GP\,0926 \citep{Hayes13,Hayes14}, and has recently been suggested for 
other GPs based on the analysis of two-dimensional COS spectra 
\citep{Yang17sizes}. 
The mechanism of the double-peak formation in the halos 
remains to be clarified, 
too. In the shell models, only a static or slowly expanding medium leads to 
the double-peaked \lya, therefore the role of kinematics must be assessed 
in the full simulation \citep[see][]{Verhamme15b}. 
Furthermore, the \lya\ halo emission can have an
in-situ origin, either from cooling radiation or UV background fluorescence, 
rather than scattering \citep{Cantalupo12,Rosdahl12,Dijkstra14}. 
If halos play a role in the \lya\ profiles, then the encountered differences 
between LIS line and shell model velocities are unsurprising. 

One could argue that high-$z$ galaxy spectra are typically 
spatially unresolved, and yet unlike in GPs  
their majority are single-peaked, 
while the incidence of multiple peaks is only 
$\sim$\,30\% \citep{Kulas12,Trainor15}. 
However, the high-$z$ results are affected by the 
increasingly neutral IGM, which removes the 
blue part of the profile \citep{Laursen11}, 
and
also by a typically low spectral resolution that blends the two peaks into one 
\citep[e.g.][]{Kulas12, Erb14}. Higher resolution data should alleviate 
at least one part of the problem. The connection 
between the blue peak and the \lya\ halo is one of the tasks to be explored 
by both numerical simulations and observations. 
The galaxy size enters the problem in yet another aspect: 
the size of the \lya\ photon source. 
The models that we used here assumed a point source. In contrast, 
the HST/COS acquisition images of GPs show a multi-knot star-formation 
structure in the near-UV. 
A similar structure will probably be reflected in the \hii\
regions, and therefore we need to ask how would this distribution
affect the radiative transfer results. 
This effect has not been addressed in the homogeneous shell models and 
is a task for further modelling.

\subsection{Clumpy shells and \lya\ produced in outflows}
\label{sec_clumpy}

Clumpy shell models 
\citep{Laursen13,Duval14,Verhamme15,Gronke16,Gronke16b,Gronke17}
have demonstrated that the \lya\ spectral profiles can be more 
complex than those from homogeneous shells. In particular, 
clumpy outflow geometries presented in \citet{Gronke16} may provide
an interesting alternative to the homogeneous shells, potentially solving 
several problems encountered in GP \lya\ profile fitting 
\citep[as was also invoked by][]{Yang16,Yang17}.

First, in clumpy outflow models, the
double-peaked \lya\ profiles are not confined to low expansion velocities. 
This is a convenient property that would remove the conflict with the  
LIS line velocities and redshifts. 
Second, some of the clumpy models can achieve a redshifted central trough
\citep[Fig.\,8 of][]{Gronke16} -- a property that is unachievable in 
homogeneous shells but is observed in some GPs. 
Third, the clumpy models may also remove the 
problem of the too broad  intrinsic \lya, as shown by some of the results
in \citet{Gronke16}. 
However, this may not be due to the clumpy nature of the model ISM,  
but rather due to the authors' assumption that \lya\ is formed inside the outflow, 
instead of a point source assumed in our models. 
More theoretical work needs to be done on the clumpy models
to assess the blue and red peak locations and their asymmetries, as well as 
connections to the \lya\ and LyC escape. A recent observation 
of a lensed $z=2.4$ galaxy reported a triple-peak \lya\ spectrum, 
similar to clumpy models, and was    
interpreted as a possible LyC leakage signature \citep{Rivera17}.

Nevertheless, we have designed a test to probe the usefulness of clumpy models
for our GPs. \citet{Gronke16} measured the residual flux in the 
modelled \lya\ central trough $F_\mathrm{trough}$, 
and found that the ratio of $F_\mathrm{trough}$ and the mean of the peak 
maxima, $\bar{F}_\mathrm{peak}$, had a different distribution in 
the clumpy models than in the homogeneous shells (their Fig.\,2). 
The ratio can span a wide range of values in the clumpy models, 
$F_\mathrm{trough}/\bar{F}_\mathrm{peak}\!\sim\!0-8,$ 
with the largest concentration around 
$F_\mathrm{trough}/\bar{F}_\mathrm{peak}\!\sim\!0.3.$ 
In contrast, $F_\mathrm{trough}/\bar{F}_\mathrm{peaks}\!<\!0.1$ 
for most of the homogeneous shell models. 
For comparison, our Fig.\,\ref{fig_trough_sep} shows that the observed 
GP $F_\mathrm{trough}/\bar{F}_\mathrm{peaks}\sim 0-0.3.$ 
These values could be explained by either model, since a small number of 
homogeneous models reach as far as 
$F_\mathrm{trough}/\bar{F}_\mathrm{peaks}\sim0.6.$
In addition, the observational data may also be affected by instrumental 
resolution, which would make the observed $F_\mathrm{trough}$ artificially high.
However, besides the interval span, 
the observed GP data inversely correlate with the \lya\ double-peak separation
(Fig.\,\ref{fig_trough_sep}). 
No such correlation was found in the clumpy models; the 
models randomly filled all of the plane.  
Conversely, homogeneous shells showed a large scatter of peak 
separations ($v_\mathrm{sep}\!\sim\!0-2000$\,\kms) 
at $F_\mathrm{trough}/\bar{F}_\mathrm{peaks} \sim0$, 
but with increasing $F_\mathrm{trough}/\bar{F}_\mathrm{peaks}$ 
the maximum separation decreases to $v_\mathrm{sep}\!<$\,500\,\kms\
at  $F_\mathrm{trough}/\bar{F}_\mathrm{peaks}\!\sim\!0.3$.
The trend in homogeneous shells thus resembles the observed 
GP data (while bearing in mind that the sample size is limited). 
In homogeneous shells, both $F_\mathrm{trough}$ and the peak separation 
are governed by \nh, therefore a correlation between them 
is expected, unlike in the case of the clumpy models 
where \lya\ escape is regulated by the properties and the 
distribution of the clumps. 

In the view of these results, the role of the clumpy models is still unclear.
A similarly hesitant conclusion was drawn by \citet{Gronke16}. 
The clumps are worth considering as they offer a multitude of possibilites 
\citep[see also][]{Gronke16b,Gronke17}. 
Model fitting of observational profiles, using various flavours 
of the clumpy models, with and without constraining their parameters, 
remains to be done. 
Tests similar to our Fig.\,\ref{fig_trough_sep} could complement the spectral
fitting.

\subsection{Symmetries of the observed and modelled \lya\ profiles}
\label{sec_sym_obsmod}

\begin{figure}[b!]
\begin{tabular}{ll}
\includegraphics[width=0.43\textwidth]{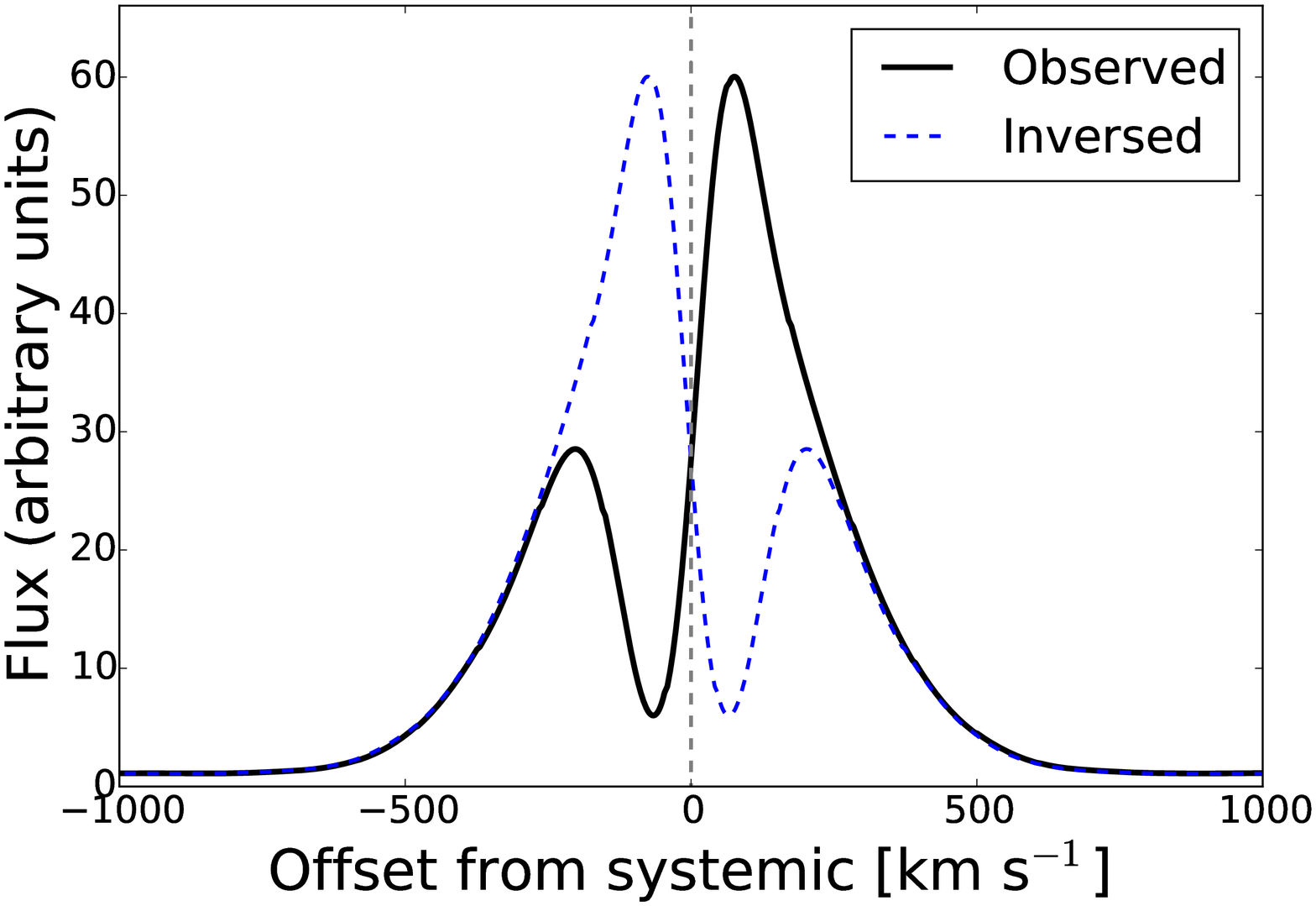}
\\
\end{tabular}
\caption{Model \lya\ profile symmetries. The solid 
black line shows the shell model, the dashed blue line its inverse with respect
to the systemic redshift. The shell model parameters:  
\nh\,$=\!10^{18}$\,cm$^{-2}$, \vexp\,$=50$\,\kms, $\tau_\mathrm{d}\!=\!1$, 
$b\!=\!20$\,\kms,
\fwhmlyai\,$=\!500$\,\kms, \ewlyai\,$=\!100$\,\kms.
}
\label{fig_inversemodel}
\end{figure}

\begin{figure*}
\begin{tabular}{lll}
\includegraphics[width=0.47\textwidth]{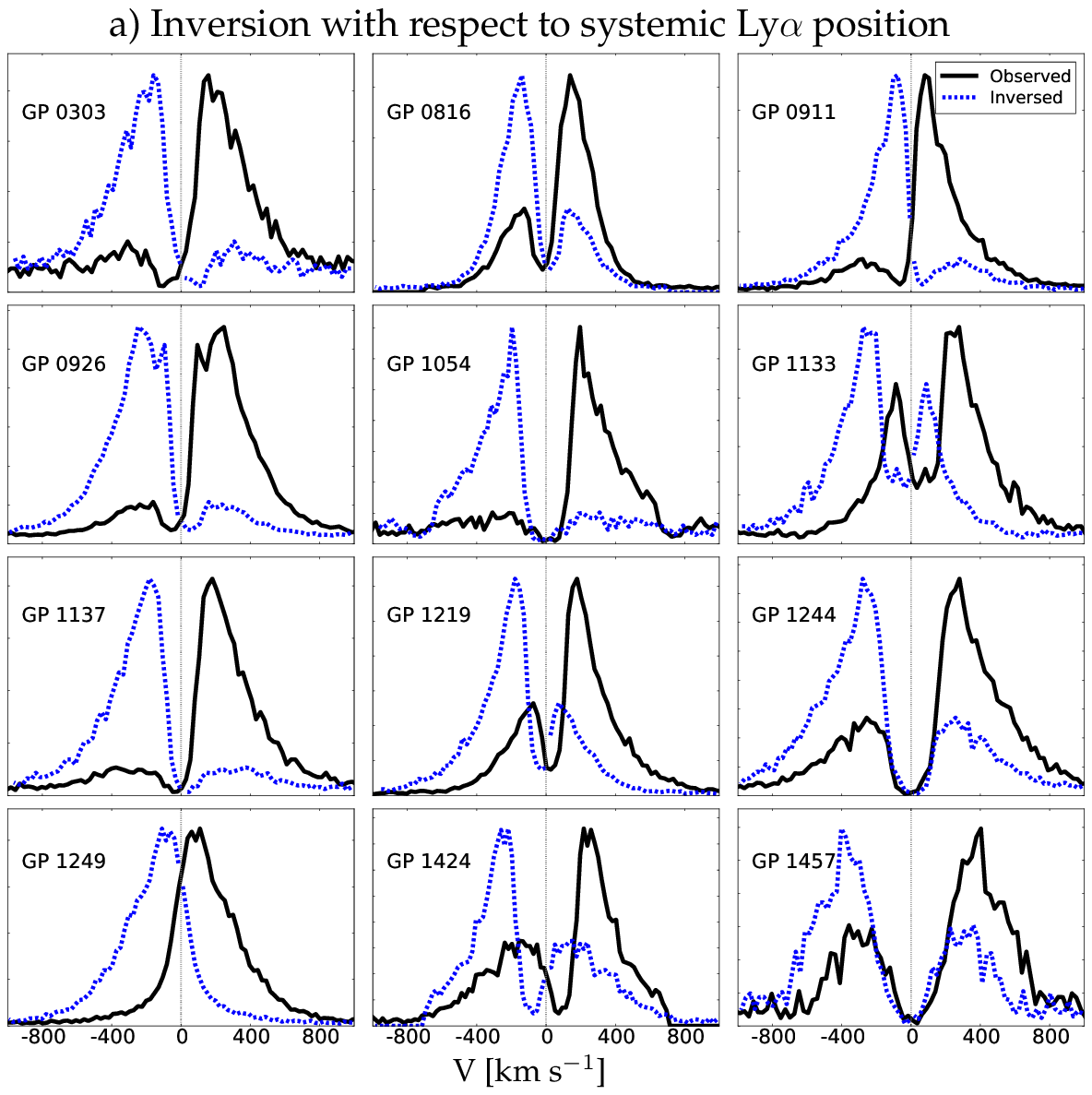}
&
&
\includegraphics[width=0.47\textwidth]{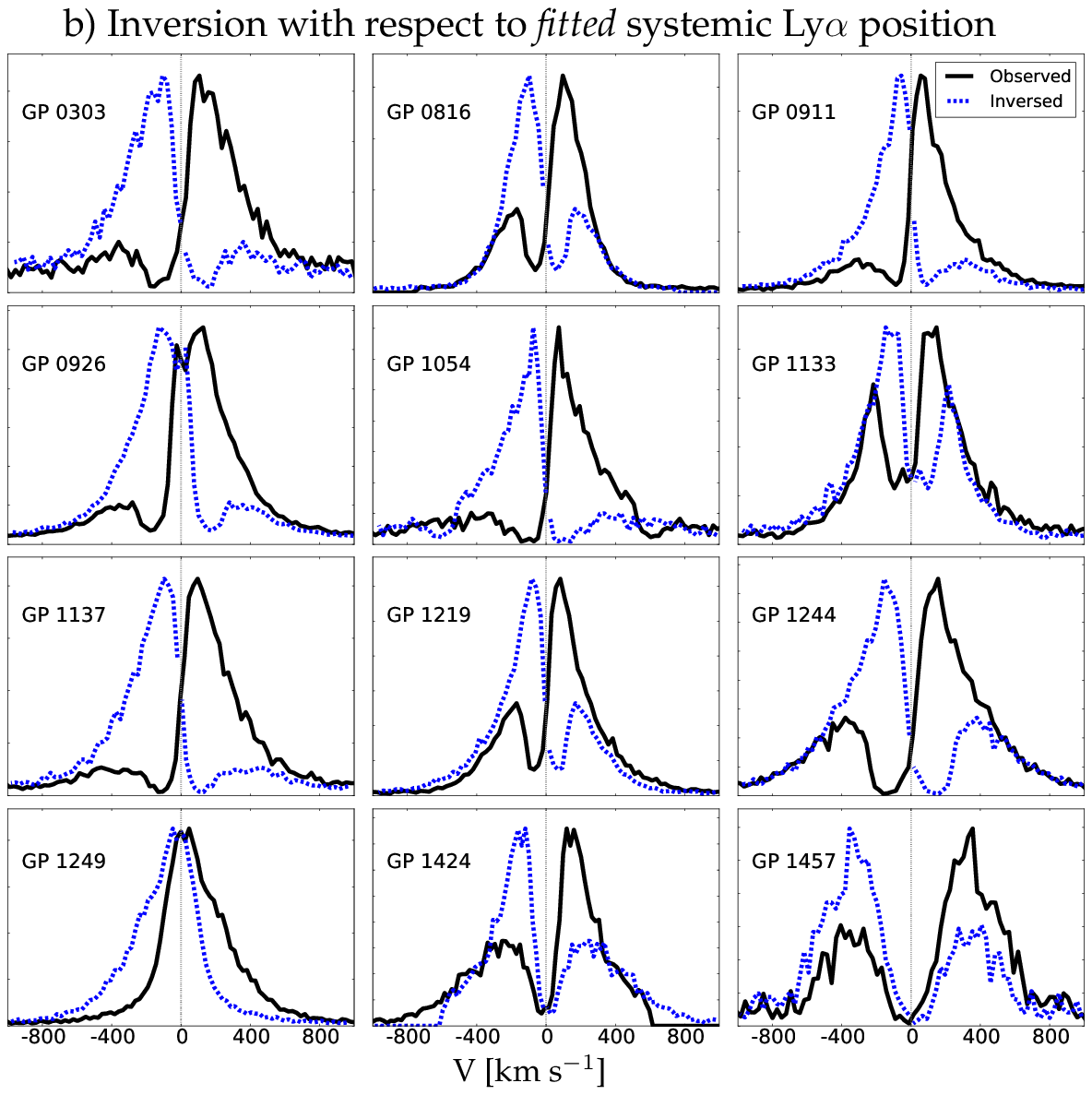}
\\
\end{tabular}
\caption{Observed \lya\ profile symmetries: 
a) with respect to the systemic velocity derived from SDSS redshift, 
b) with respect to the systemic velocity derived from best-fitting model. 
The solid black line shows the observed data, the blue line their inversion. 
The red and blue wings are more symmetric in case (b) than in case (a).
}
\label{fig_inverse}
\end{figure*}

We here explore symmetries of the shell model \lya\ profiles 
and compare them with those observed in the GPs. 
Symmetries in peak positions, wings, and troughs
could provide additional 
insight into the incompatibilities between the GP data and shell models or
their parameters.      

It was noted by \citet{Henry15} that 
it is mainly the blue-peak position that 
correlates with the GP double-peak separation, 
and thus with the \lya\ and LyC escape \citep{Verhamme15,Verhamme17}.
\citet{Yang17} then showed that both peaks correlate 
with \fesclya\ in a larger GP sample. We here test if the same
is true for the models. We work with twelve GPs, 
therefore we have to bear in mind the effects of the limited sample size. 
We presented the GP asymmetric blue and red \lya\ peak 
positions in Fig.\,\ref{fig_symm}. We here study the \lya\ 
double-peak separation versus the blue and red peak positions 
measured from the systemic redshift both in the observed and modelled 
spectra (Fig.\,\ref{fig_symm2}a,b). 
The double-peak separation is a proxy for the \lya\ and LyC escape, and 
we explore how the blue and red peaks correlate with it and what symmetries
exist in the observations and the models.  
We include the shell models that were among the twenty best fits for each 
target (see Section\,\ref{sec_unconstrained}). We find stronger correlations 
for the blue peak both in the COS data and in the models. 
However, a clear shift between the red and blue peak positions exists in the 
models, not seen in the observations.  
Models with strong blue peaks (blue-to-red flux ratio $>0.3,$ 
Table\,\ref{tab_ewbr}) have the most symmetric peak positions 
(i.e. red and blue squares plotted close to each other 
in Fig.\,\ref{fig_symm2}b). Conversely, the weak blue peak models are 
responsible for the scatter in that plot.             
On the other hand, if we measure the peak positions from the central trough, 
instead of the systemic velocity, then we obtain plots of 
Fig.\,\ref{fig_symm2}c,d.
Both the observed and modelled spectra show more symmetry with respect to the 
trough than to the systemic velocity. 
The correlation with the peak separation is strong for both the
blue and red peaks and for both the observed and modelled spectra in this 
case. This result is another illustration of 
why our models failed to reproduce the observed 
spectra with applied constraints, while 
they were usable in the modelling with free redshift; they attained the required symmetry with the modified redshift. 
Using the modified redshift, the blue and red peak positions would resemble the 
models in Fig.\,\ref{fig_symm2}b.

We have further tested the symmetry of the \lya\ line wings
by exploring the \lya\ profiles together with their plots 
flipped with respect to the systemic velocity 
(Figs.\,\ref{fig_inversemodel} and \ref{fig_inverse}). 
The modelled double-peaks have symmetric wings with 
respect to the systemic redshift (Fig.\,\ref{fig_inversemodel})  
but with the condition of a broad \fwhmlyai, 
equivalent to the one that fits the observed GP profiles. 
The wing symmetry is 
created by small effects of radiative transfer far from the line centre, and  
by a symmetric intrinsic profile.  
On the other hand, models with a narrow \fwhmlyai\ 
(as in Fig.\,\ref{fig_constrained})
have red wings stronger than the blue ones, 
due to radiative transfer effects.   
For the observed GP spectra, the red wing is also mostly stronger 
than the blue one (Fig.\,\ref{fig_inverse}a), 
despite the fact that the wings are as broad as in the model of 
Fig.\,\ref{fig_inversemodel}.
On the contrary, if we consider $z$ derived from the
\lya\ fits rather than SDSS, the observed wings 
become symmetric (Fig.\,\ref{fig_inverse}b).

We conclude that the free fitting process modifies $z$ 
to obtain spectra that are more symmetric in the red and blue peak position 
with respect to the systemic redshift. The new symmetry makes the data 
compatible with the shell models.  
As a consequence, the line wings become symmetric. The wing symmetry 
of the model spectra is in turn achievable by assuming 
a large intrinsic \fwhmlyai.
A comparison with clumpy models and other geometries would be useful 
to assess how unique is the symmetry resemblance between the data and 
the shifted models, resulting in the alignment of the trough position and wing 
shape, and in the possibility of finding models that fit the double peaks. 
This exercise still does not clarify the reasons for the discrepancy in $z.$ 
It does not answer the question of how  appropriate the shell models are,  
or if the resemblance 
between the unconstrained fitting models and the data is a coincidence.  
We showed in Section\,\ref{sec_ewbr} that the discrepancies in individual 
fitting parameters were tied to the spectral shape, namely to the blue 
peak $EW$. We also showed that the blue peaks were related to 
\lya\ and LyC escape.
In this light, the resemblance between symmetries of modelled and 
observed (shifted) spectra appears surprising and requires   
more theoretical work.  

\subsection{\lya\ sources}

Our models considered recombination as the only source of \lya. 
Some of the fitting parameter discrepancies, namely in $z$ and \fwhmlyai, 
were evaluated by comparing the \lya\ and \hb\ lines under the assumption that 
the same recombination process was the origin of both lines. 
However, other \lya\ production mechanisms are possible and could 
be responsible for part of the fitting problems. 

Collisional excitation is one such process that could cause 
kinematic differences between \lya\ and \hb. 
Collisional excitation affects more the first excited level than the 
higher energy levels, 
leading to the \lya/\ha\ emissivity ratio $\sim\!\!100$
\citep{Dijkstra14}, that is, an order of magnitude higher than in the case of 
recombination ($\sim\!\!8$). 
In violent conditions inside GPs, characterized by large star-formation 
rates and high excitation, strong collisional processes can be expected
\citep{Oti12}. Typical GP electron temperatures are relatively high, 
$\sim\!\!15\,000$\,K, favourable to the collisional excitation scenario  
\citep{Jaskot13}. 
Collisional contribution could explain the intrinsic \lya\ profile 
broadening and the redshift discrepancy between the 
modelled intrinsic \lya\ and the observed \hb.
We have also previously mentioned the possibility of \lya\ production 
in outflowing medium (Sect.\,\ref{sec_clumpy}), 
which could be due to a number of different processes;
their impact on the resulting spectrum needs to be further explored. 

Other possible sources of \lya\ emission include fluorescence 
\citep{Cantalupo12,Cantalupo14} and gravitational cooling 
\citep{Dijkstra06,Dijkstra14}. Both processes, acting in the outer 
layers of the ISM, would be able to produce a large \fwhmlyai\ 
\citep{Hashimoto15}.
In addition, Fermi-like acceleration of \lya\ photons across shock fronts
was suggested as an alternative origin of the \lya\ blue peaks \citep{Chung16,
Neufeld88}. Tests for these
predictions are still missing. \citet{Martin15} searched for the signatures of 
Fermi acceleration in ULIRGs and concluded that no compelling evidence was found,
but they admitted that the process can play a role in some objects.

Finally, we note that a detailed exploration of the recombination sources 
alone would also be
useful. The SDSS \hb\ spectra that we used did not have sufficient resolving 
power ($R\!\sim\!2000$) to show the complete emission line structure. 
\citet{Amorin12} observed several green peas with a high-resolution 
echelle spectrograph ($R\!>\!\!10\,000$), and found  
complex \ha\ profiles with several distinct kinematic components. If 
the components come
from different regions of the galaxy, conditions for the \lya\ transfer in 
each of them can be different and could thus affect differently 
the resulting \lya\ profile. 
The sample of \citet{Amorin12} unfortunately does not overlap
with our \lya\ sample and thus could not be tested.

\section{Summary and conclusions}

We have studied in detail the \lya\ spectra of twelve green pea
galaxies, which are an unusual population of 
strongly star-forming compact 
galaxies at $z\sim0.2,$ and which resemble high-redshift \lya\ emitters 
in their mass, metallicity, star-formation rate, and possibly 
ionizing continuum leakage. Eleven out of the twelve studied 
GPs have double-peaked emission-line \lya\ spectral profiles.
The spectra show no signs of broad underlying \lya\ absorption 
(which is often observed in low-$z$ star-forming galaxies), 
with two weak exceptions.   
Furthermore, they have non-zero residual flux in the central trough 
that separates the blue and red peaks. Together with a small peak separation, 
these properties indicate low \hi\ column densities, 
based on the criteria of \citet{Verhamme15}.
The central \lya\ trough is redshifted from the systemic velocity
in several GPs, which is unusual in the context of the known observations and 
models.

We applied the {\tt MCLya} Monte Carlo \lya\ 
radiative transfer code \citep{Verhamme06}, which uses the geometry of 
 expanding homogeneous shells, to fit the 
GP \lya\ spectra and derive their ISM parameters. 
In the first step, we applied detailed constraints on the  
fitting parameters, inferred from ancillary UV and optical spectra. 
The models did not correctly reproduce the observed GP \lya\ spectra in this
case (Fig.\,\ref{fig_constrained}). 
We thus proceeded to unconstrained fitting 
\citep[similar to][]{Yang16,Yang17}, which correctly
reproduced the spectral profiles (Fig.\,\ref{fig_fits}), 
but the best-fitting model parameters 
were in disagreement with the ancillary data (Fig.\,\ref{fig_parameters}).
In particular: 1) The redshifts derived from the 
\lya\ fitting were in all cases
larger than those from the SDSS optical emission lines 
($\Delta z\!=\!10\!-\!250$\,\kms); 
2) The best-fit outflow velocities were typically $\lesssim\!150$\,\kms, 
whereas the UV LIS line velocities were distributed in 
the interval $0-300$\,\kms; 
3) The modelled \fwhmlyai\ of the intrinsic
\lya\ line was a factor of two to four larger 
than the measured \fwhmhb\ in each target. 
We found a link between the fit parameter discrepancies 
and the double-peak character of the \lya\ profiles, 
namely the \ewlyab/\ewlyar\ ratio of the blue and red peak equivalent widths 
(Fig.\,\ref{fig_ewbr}). 
 
We propose two interpretations for the data-model disagreement. 
First, the ancillary data may be 
inappropriate to constrain the models; the LIS lines may
not probe the environment where the \lya\ transfer takes place and the 
intrinsic \lya\ may not be produced by the same mechanism as \hb. 
We showed that by modifying $z,$ the observed 
\lya\ trough positions would become compatible with the models 
(Fig.\,\ref{fig_ewbr}b) and the observed \lya\ profile symmetries would 
correspond to those of the homogeneous shell models 
(Figs.\,\ref{fig_symm2}-\ref{fig_inverse}). 
Second, the blue \lya\ peaks of GPs may not originate in a static ISM, 
as is the case in the homogeneous shell models, or 
at least not in the gas probed by the UV LIS lines. The blue peak formation
mechanisms, either at a source or by the radiative transfer, need to be further 
investigated. The \ewlyab/\ewlyar\ ratio correlates 
with the \lya\ escape fraction, and with the \lya\ peak separation, 
which suggests that the GP blue peaks are associated with low \nh, and with 
the \lya\ and LyC escape, rather than kinematics. 
A connection between the blue peak position 
and \nh\ was
previously proposed in the literature \citep{Henry15,Yang16,Yang17}, 
while we here extend this effect to the blue peak flux.
 
We considered alternative models to reproduce the GP \lya\ profiles. 
No combination of outflowing and inflowing homogeneous shells was found to be 
compatible with the observed GP spectra. 
Clumpy models such as those of \citet{Gronke16} are
a promising option, as they produce double peaks of various shapes 
by other mechanisms than kinematics. However, more theoretical work is needed
to check their compatibility with observations. We could not confirm the 
compatibility based on the measured residual flux in the central trough of
\lya\ spectra. We found that it correlates with the red-blue \lya\ 
peak separation 
in GPs, which is a trend expected in the homogeneous case and not in 
the clumpy models \citep{Gronke16}.  
Nevertheless, the small observational sample may only cover a portion of the 
parameter space and a larger sample will be needed to extend this study. Also,  
various versions of the clumpy models may provide different results.

For future work, 
high-resolution \ha\ or \hb\ spectra would be useful to  
provide more details about the \hii\ regions kinematics, that is, about the 
\lya\ source. If multiple kinematic components are present, their 
impact on the resulting \lya\ profile needs to be explored in the models.  
Possible contributions of non-recombination processes 
to the \lya\ spectra need to be tested. 
Finally, model fitting with the use of clumpy geometries 
should clarify whether the clumps are a 
solution to the problem of mismatching ISM parameters.

\begin{acknowledgements}
We thank the anonymous referee for improving the clarity of the paper. 
We thank Matthew Hayes for providing the fitting tool essential to 
the \lya\ modelling. 
I.O. appreciated the discussions on stellar atmosphere models
with Ji\v r{\'\i} Krti\v cka, and 
acknowledges the support from the Czech Science Foundation grant 
number 17-06217Y. 
This work is based on data from HST-GO-13293, which contributed support to MSO and AEJ, and archive data from HST-GO-12928 and 11727. 
The data were obtained from the Mikulski Archive for Space Telescopes (MAST). 
STScI is operated by the Association of Universities for Research in Astronomy, Inc., 
under NASA contract NAS5-26555. Support for MAST for non-HST data is provided by the 
NASA Office of Space Science via grant NNX09AF08G and by other grants and contracts.
We have made use of SDSS data.
Funding for SDSS-III has been provided by the Alfred P. Sloan Foundation, the Participating Institutions, the National Science Foundation, and the U.S. Department of Energy Office of Science. The SDSS-III web site is http://www.sdss3.org/.
SDSS-III is managed by the Astrophysical Research Consortium for the Participating Institutions of the SDSS-III Collaboration including the University of Arizona, the Brazilian Participation Group, Brookhaven National Laboratory, Carnegie Mellon University, University of Florida, the French Participation Group, the German Participation Group, Harvard University, the Instituto de Astrofisica de Canarias, the Michigan State/Notre Dame/JINA Participation Group, Johns Hopkins University, Lawrence Berkeley National Laboratory, Max Planck Institute for Astrophysics, Max Planck Institute for Extraterrestrial Physics, New Mexico State University, New York University, Ohio State University, Pennsylvania State University, University of Portsmouth, Princeton University, the Spanish Participation Group, University of Tokyo, University of Utah, Vanderbilt University, University of Virginia, University of Washington, and Yale University.
We have made use of the ADS and NED databases.  
\end{acknowledgements}

\bibliographystyle{aa} 
\bibliography{references} 

%
%

\end{document}